\documentclass[iop]{emulateapj}
\usepackage{natbib,amsmath,graphicx,longtable,amssymb,latexsym,epsf,sidecap}
\begin{document}

\title{Quasars Probing Quasars IX. The Kinematics of the Circumgalactic Medium Surrounding
$z\sim2$ Quasars}

\author{Marie Wingyee Lau\altaffilmark{1,2}, J. Xavier Prochaska\altaffilmark{2},
Joseph F. Hennawi\altaffilmark{3}
}
\altaffiltext{1}{Email: lwymarie@ucolick.org}
\altaffiltext{2}{Department of Astronomy and Astrophysics, UCO/Lick Observatory, University of
California, 1156 High Street, Santa Cruz, CA 95064}
\altaffiltext{3}{Department of Physics, Broida Hall, University of California, Santa Barbara,
CA 93106-9530}

\begin{abstract}
We examine the kinematics of the gas in the environments of galaxies hosting quasars at $z\sim2$.
We employ 148 projected quasar pairs to study the circumgalactic gas of the foreground quasars in
absorption. The sample selects foreground quasars with precise redshift measurements, using
emission-lines with precision $\lesssim300\,{\rm km\,s^{-1}}$ and average offsets from the
systemic redshift $\lesssim|100\,{\rm km\,s^{-1}}|$. We stack the background quasar spectra at the
foreground quasar's systemic redshift to study the mean absorption in \ion{C}{2}, \ion{C}{4},
and \ion{Mg}{2}.
We find that the mean absorptions exhibit large velocity widths
$\sigma_v\approx300\,{\rm km\,s^{-1}}$. Further, the mean absorptions appear to be asymmetric about
the systemic redshifts. The mean absorption centroids exhibit small redshift relative to the
systemic $\delta v\approx+200\,{\rm km\,s^{-1}}$, with large intrinsic scatter in the centroid
velocities of the individual absorption systems.
We find the observed widths are consistent with gas in gravitational
motion and Hubble flow. However, while the observation of large widths alone does not require
galactic-scale outflows, the observed offsets suggest that the gas is on average outflowing from the
galaxy. The observed offsets also suggest that the ionizing radiation from the foreground quasars
is anisotropic and/or intermittent.
\end{abstract}


\section{Introduction}
\label{sec:introduction}

Galaxy formation and evolution are driven by the flows of gas into and out of their interstellar
medium. Current theories demand that star-forming galaxies maintain these flows. Gas accretes,
cools, and adds to the fuel supply, while star formation feedback heats gas, blows it out of
galaxies, and regulates star formation \citep[for a review see][]{SomervilleDave15}.

Direct observations of galactic flows are difficult to acquire. Detecting the presence of
the gas is itself challenging. Either the gas mass is too small, or the gas density is too low for
the detection of line-emission, e.g.\ 21\,cm, Ly$\alpha$, or H$\alpha$ from \ion{H}{1}. Resolving
the kinematics and establishing the mass flux pose an even greater challenge. These challenges are
accentuated for distant, young galaxies, where flows of gas are predicted to prevail
\citep{Keres+09,Fumagalli+11}. Therefore, with rare exceptions,
\citep[e.g.,][]{Cantalupo+14,Hennawi+15}, the community has relied on absorption-line spectroscopy
to detect and characterize the gas surrounding galaxies
\citep[e.g.,][]{BergeronBoisse91,Steidel+10,Prochaska+11,Tumlinson+13}.
In a previous paper of the Qussars Probing Quasars series \citep[][, hereafter QPQ8]{QPQ8}, we
measured the velocity field for \ion{C}{2}\,1334 and \ion{C}{4}\,1548, finding that the
circumgalactic medium frequently exhibits large velocity widths that are offset from the systemic
redshift.
From a sample of 7 \ion{C}{2} systems and 10 \ion{C}{4}
systems, we measured the velocity interval that encompasses 90\% of the total optical depth,
$\Delta v_{90}$, and the $1\sigma$ dispersion relative to the profile centroid, $\sigma_v$.
The median $\Delta v_{90}$ is $555\,{\rm km\,s^{-1}}$ for \ion{C}{2}\,1334 and is
$342\,{\rm km\,s^{-1}}$ for \ion{C}{4}\,1548. The median $\sigma_v$ is $249\,{\rm km\,s^{-1}}$ for
\ion{C}{2}\,1334 and is $495\,{\rm km\,s^{-1}}$ for \ion{C}{4}\,1548. These velocity fields exceed
all previous measurements from galaxies and/or absorption systems at any epoch.
The offsets, $\delta v$, are often positive, with the sign convention that positive
velocities indicate a redshift from the systemic.

With absorption-line spectroscopy of background sightlines, other researchers also have had
success in characterizing the flows of gas around galaxies. \cite{Rakic+12} found a net
large-scale inflow around star-forming galaxies, or a Kaiser effect for gas on
1\textendash2\,Mpc scales. \cite{Ho+17} found gas spiraling inward near the disk plane of
star-forming galaxies on $<100$\,kpc scales. \cite{Johnson+15} studied the CGM surrounding
$z\sim1$ quasars. They found large peculiar motions in the gas exceeding the expected virial
velocity, with quasar-driven outflows being one possible explanation. However, we consider that
their velocity spreads can be better quantified. Other
existing studies that found large velocity spreads are single sightlines
\citep[e.g.,][]{Tripp+11,RudieNewmanMurphy17}, or with gas tracing
a higher ionization state than the QPQ absorption systems \citep[e.g.,][]{Churchill+12}, or where
the average velocity spread is smaller than that measured in QPQ8. In \cite{Gauthier13}, where a
single sightline is reported, the $\Delta v_{90}$ of the \ion{Mg}{2} absorption is less than the
average of the QPQ8 \ion{C}{2} absorption. In \cite{Muzahid+15}, an absorber is found with
$\Delta v_{90}$ smaller than $555\,{\rm km\,s^{-1}}$ in \ion{O}{6} and \ion{N}{5}, and still
smaller for other ions. The \cite{Zahedy+16} sample likewise has average $\Delta v_{90}$ smaller
than that of QPQ8.

A significant limitation of absorption-line analysis of transverse sightlines, especially
regarding galactic-scale flows, is the inherent symmetry of the experiment. One generally lacks
any constraint on the distance of the gas along the sightline. Positive or negative velocities
with respect to the galaxy may be interpreted as gas flowing either toward or away from the
system. ``Down-the-barrel'' observations break this symmetry, and have generally provided evidence
for flows away from galaxies \citep{Rupke+05,Martin05,Weiner+09,Rubin+14}. However, these data are
frequently at low spectral resolution which limits one's sensitivity to inflowing gas.

In this paper (hereafter QPQ9), we examine the flows of gas in the environments of massive
galaxies hosting quasars. Our approach leverages a large dataset of quasar pairs
\citep[][hereafter QPQ1]{QPQ1} to use the standard techniques of absorption-line spectroscopy
with background quasars. These quasar pairs
have angular separations that correspond to less than 300\,kpc projected separation at the
foreground quasar's redshift. Our previous publications from these quasar pairs have established
that these galaxies are surrounded by a massive, cool, and enriched CGM
\citep[QPQ5, QPQ6, QPQ7:][]{QPQ5,QPQ6,QPQ7}. We have collected a sample of 148 background
spectra that are paired with foreground quasars with precisely measured redshifts. Among the
sightlines in the QPQ9 sample, 13 have spectral resolution $R>5000$ from echellette or echelle
observations, and have been analyzed separately in QPQ8 (see their Figure 8, 9, and 10, and their
Appendix). In QPQ8, where the individual metal-bearing absorption components are resolved, 15 out
of the 21 components are at positive velocities relative to the systemic redshift. For this
current work, we stack spectra of all resolutions instead of performing a component-by-component
analysis. Our primary scientific interests are twofold: (i) search for signatures of galactic-scale
outflows from the central galaxy, presumably driven by recent star formation and/or active galactic
nuclei feedback; (ii) characterize the dynamics of the gas around these massive systems. We further
describe an aspect of this experiment that offers a unique opportunity to study galactic-scale
flows: we argue that, the anisotropic or intermittent radiation from the foreground quasars may
break the symmetry in the velocity field of circumgalactic absorbers. If the ionizing radiation
field is asymmetric, the absorbers may also distribute asymmetrically. Alternatively, finite quasar
lifetime will result in different radiation fields impinged on the gas closer to versus further
away from the background quasar, due to different light travel times. \cite{KirkmanTytler08}
reported an asymmetry in \ion{H}{1} absorption on scales larger than the CGM, and gave similar
arguments on anisotropy or intermittence.

We adopt a $\Lambda$CDM cosmology with $\Omega_M=0.26, \Omega_\Lambda=0.74$, and
$H_0=70\,{\rm km\,s^{-1}\,Mpc^{-1}}$. Distances are proper unless otherwise stated. When referring
to comoving distances we include explicitly an $h^{-1}$ term and follow modern convention of
scaling to the Hubble constant $h = H_0/(70\,{\rm km\,s^{-1}\,Mpc^{-1}})$.

\section{The Experiment}
\label{sec:experiment}

\LongTables
\begin{deluxetable*}{ccccccccc}
\tablewidth{0pc}
\tablecaption{Properties of the Projected Quasar Pairs in the QPQ9 sample\label{tab:sample}}
\tabletypesize{\scriptsize}
\setlength{\tabcolsep}{0in}
\tablehead{\colhead{Foreground Quasar} & \colhead{$z_{\rm fg}$} & 
\colhead{Line for $z_{\rm fg}^a$} & \colhead{Background Quasar} & 
\colhead{$z_{\rm bg}$} & \colhead{BG Quasar Instrument} & 
\colhead{$R_\perp$ (kpc)} & \colhead{$g_{\rm UV}$}} 
\startdata 
J003423.05$-$1050020 & 1.8388 & MgII & J003423.44-104956.3 & 1.948 & LRIS & 67 & 11938 \\ 
J004220.66+003218.7 & 1.9259 & MgII & J004218.72+003237.1 & 3.048 & BOSS & 299 & 97 \\ 
J004745.49+310120.3 & 1.9706 & MgII & J004745.61+310138.3 & 2.695 & BOSS & 157 & 1723 \\ 
J004757.26+144741.0 & 1.6191 & MgII & J004757.88+144744.7 & 2.757 & BOSS & 82 & 6697 \\ 
J005717.36$-$000113.3 & 2.1611 & [OIII] & J005718.99-000134.7 & 2.511 & BOSS & 271 & 1283 \\ 
J010323.84$-$000254.2 & 1.7506 & MgII & J010324.37-000251.3 & 2.306 & BOSS & 74 & 5462 \\ 
J014328.77+295436.8 & 1.8007 & MgII & J014330.89+295439.9 & 2.018 & BOSS & 243 & 997 \\ 
J014917.11$-$002141.6 & 1.6834 & MgII & J014917.46-002158.5 & 2.159 & SDSS & 155 & 2390 \\ 
J021416.96$-$005229.1 & 1.8002 & MgII & J021416.12-005251.5 & 2.332 & BOSS & 225 & 564 \\ 
J022447.89$-$004700.4 & 1.6959 & MgII & J022448.85-004638.9 & 2.188 & BOSS & 226 & 293 \\ 
J023018.27$-$033319.4 & 2.3817 & MgII & J023019.99-033315.0 & 2.985 & BOSS & 221 & 1444 \\ 
J023315.44$-$000303.6 & 1.7205 & MgII & J023315.75-000231.4 & 1.839 & BOSS & 286 & 101 \\ 
J023946.43$-$010640.4 & 2.299 & [OIII] & J023946.45-010644.1 & 3.124 & BOSS & 32 & 20931 \\ 
J024603.68$-$003211.8 & 1.603 & MgII & J024602.35-003221.6 & 2.153 & BOSS & 195 & 1889 \\ 
J025038.68$-$004739.2 & 1.8538 & MgII & J025039.82-004749.6 & 2.445 & BOSS & 175 & 4007 \\ 
J034138.16+000002.9 & 2.1246 & MgII & J034139.19-000012.7 & 2.243 & GMOS-N & 190 & 392 \\ 
J040955.87$-$041126.9 & 1.7166 & MgII & J040954.21-041137.1 & 2.0 & SDSS & 235 & 715 \\ 
J072739.55+392855.3 & 1.9853 & MgII & J072739.72+392919.5 & 2.433 & BOSS & 210 & 403 \\ 
J075009.25+272405.2 & 1.7713 & MgII & J075008.27+272404.5 & 1.802 & LRIS & 114 & 1370 \\ 
J075259.81+401128.2 & 1.8844 & MgII & J075259.14+401118.2 & 2.121 & SDSS & 110 & 1060 \\ 
J080049.89+354249.6 & 1.9825 & [OIII] & J080048.74+354231.3 & 2.066 & LRIS & 201 & 2074 \\ 
J080537.29+472339.3 & 1.8913 & MgII & J080538.78+472404.8 & 2.964 & BOSS & 259 & 367 \\ 
J080945.17+453918.1 & 2.0392 & MgII & J080948.22+453929.0 & 2.278 & BOSS & 292 & 195 \\ 
J081223.17+262000.9 & 1.6427 & MgII & J081223.89+262012.5 & 2.17 & BOSS & 132 & 3442 \\ 
J081419.58+325018.7 & 2.1744 & [OIII] & J081420.38+325016.1 & 2.213 & GMOS-N & 90 & 2899 \\ 
J081832.87+123219.9 & 1.7032 & MgII & J081833.97+123215.4 & 2.234 & BOSS & 147 & 6609 \\ 
J082346.05+532527.8 & 1.6467 & MgII & J082347.49+532519.1 & 1.86 & BOSS & 136 & 820 \\ 
J082421.01+531249.3 & 2.0855 & MgII & J082420.02+531315.2 & 2.165 & BOSS & 237 & 340 \\ 
J082843.37+454517.3 & 1.873 & MgII & J082844.87+454518.2 & 1.987 & LRIS & 137 & 525 \\ 
J083030.38+545228.8 & 1.6702 & MgII & J083029.11+545210.3 & 3.337 & BOSS & 188 & 1506 \\ 
J083713.56+363037.3 & 1.8364 & MgII & J083712.69+363037.7 & 2.301 & MODS1 & 92 & 5012 \\ 
J083757.91+383727.1 & 2.0624 & H$\alpha$ & J083757.13+383722.4 & 2.251 & LRIS & 89 & 8609 \\ 
J083854.52+462124.4 & 1.7596 & MgII & J083852.94+462137.6 & 2.163 & BOSS & 184 & 673 \\ 
J084158.47+392120.0 & 2.0414 & [OIII] & J084159.26+392139.0 & 2.214 & LRIS & 183 & 1514 \\ 
J084511.89+464135.5 & 1.6295 & MgII & J084509.64+464113.0 & 1.898 & BOSS & 283 & 135 \\ 
J085019.43+475538.5 & 1.8164 & MgII & J085021.17+475516.0 & 1.891 & BOSS & 249 & 392 \\ 
J085151.38+522901.6 & 1.9738 & MgII & J085154.53+522910.6 & 2.031 & BOSS & 262 & 265 \\ 
J085249.45+471423.1 & 1.6468 & MgII & J085248.55+471419.3 & 1.688 & BOSS & 87 & 7078 \\ 
J085358.36$-$001108.0 & 2.4016 & [OIII] & J085357.49-001106.2 & 2.579 & MagE & 112 & 1231 \\ 
J085629.48+551450.2 & 1.6228 & MgII & J085630.45+551417.5 & 1.932 & BOSS & 296 & 466 \\ 
J085737.58+390120.5 & 1.9529 & MgII & J085738.00+390136.0 & 2.848 & BOSS & 150 & 1984 \\ 
J090417.94+004148.2 & 1.6193 & MgII & J090419.12+004205.1 & 1.645 & SDSS & 214 & 1035 \\ 
J090657.78+100121.4 & 1.6965 & MgII & J090657.62+100105.6 & 2.525 & BOSS & 141 & 6960 \\ 
J091046.44+041458.5 & 2.0461 & [OIII] & J091046.69+041448.4 & 2.377 & MagE & 95 & 11897 \\ 
J091217.57+413933.5 & 1.7764 & MgII & J091215.75+413948.2 & 2.198 & BOSS & 220 & 790 \\ 
J091234.27+305616.2 & 1.6237 & MgII & J091236.32+305626.5 & 2.146 & BOSS & 247 & 514 \\ 
J091338.33$-$010708.7 & 2.7491 & MgII & J091338.97-010704.6 & 2.916 & XSHOOTER & 89 & 5830 \\ 
J091432.02+010912.4 & 2.1404 & [OIII] & J091430.85+010927.5 & 2.475 & BOSS & 199 & 1222 \\ 
J091551.72+011900.2 & 1.9706 & MgII & J091553.37+011911.4 & 2.102 & BOSS & 236 & 970 \\ 
J092405.06+474611.4 & 2.0556 & MgII & J092402.85+474600.7 & 2.098 & BOSS & 214 & 341 \\ 
J092417.65+392920.3 & 1.8864 & MgII & J092416.72+392914.6 & 2.08 & LRIS & 106 & 2788 \\ 
J092543.88+372504.9 & 2.0704 & MgII & J092544.71+372503.5 & 2.314 & BOSS & 94 & 3890 \\ 
J093226.34+092526.1 & 2.4172 & [OIII] & J093225.66+092500.2 & 2.602 & MagE & 238 & 774 \\ 
J093317.43+592027.4 & 1.8617 & MgII & J093320.57+592036.5 & 2.633 & BOSS & 224 & 1441 \\ 
J093640.35$-$005840.1 & 2.2098 & MgII & J093642.12-005831.3 & 2.731 & BOSS & 250 & 496 \\ 
J093936.83+482115.0 & 1.8878 & MgII & J093938.97+482059.4 & 2.415 & BOSS & 230 & 528 \\ 
J093952.56+505207.4 & 1.6105 & MgII & J093954.75+505148.8 & 2.476 & BOSS & 242 & 128 \\ 
J094133.64+230840.1 & 1.7762 & MgII & J094135.61+230845.8 & 2.551 & BOSS & 243 & 1700 \\ 
J094906.23+465938.4 & 1.7204 & MgII & J094906.52+465909.6 & 2.173 & BOSS & 253 & 243 \\ 
J095127.06+493248.3 & 1.7407 & MgII & J095126.22+493218.8 & 1.83 & BOSS & 268 & 293 \\ 
J095858.88+491253.1 & 2.0505 & MgII & J095858.06+491307.5 & 2.155 & BOSS & 144 & 722 \\ 
J100046.45+033708.8 & 1.7006 & MgII & J100048.52+033708.8 & 2.353 & BOSS & 271 & 255 \\ 
J100509.56+501929.8 & 1.8176 & MgII & J100507.07+501929.8 & 2.019 & LRIS & 211 & 330 \\ 
J100627.47+480420.0 & 2.3034 & [OIII] & J100627.11+480429.9 & 2.597 & BOSS & 90 & 2886 \\ 
J100913.91+023612.4 & 1.7359 & MgII & J100913.33+023643.0 & 2.216 & BOSS & 287 & 342 \\ 
J100941.35+250104.1 & 1.8703 & MgII & J100940.58+250053.9 & 1.981 & LRIS & 127 & 3588 \\ 
J101001.51+403755.5 & 2.1924 & H$\alpha$ & J101003.47+403754.9 & 2.505 & BOSS & 191 & 5421 \\ 
J101323.89+033016.0 & 1.9401 & MgII & J101322.23+033009.1 & 2.273 & BOSS & 219 & 426 \\ 
J101753.38+622653.4 & 1.6528 & MgII & J101750.44+622648.2 & 2.738 & BOSS & 184 & 922 \\ 
J101947.11+494835.8 & 1.6224 & MgII & J1019470+494849.1 & 1.652 & LRIS & 117 & 881 \\ 
J102007.23+611955.0 & 1.7909 & MgII & J102010.05+611950.3 & 2.387 & BOSS & 180 & 1817 \\ 
J102259.33+491125.8 & 1.9757 & MgII & J102259.97+491151.7 & 2.469 & BOSS & 231 & 212 \\ 
J102821.26+240121.8 & 1.8709 & MgII & J102822.18+240057.4 & 2.414 & BOSS & 240 & 834 \\ 
J103443.62+085702.0 & 1.6395 & MgII & J103442.26+085645.7 & 2.766 & BOSS & 233 & 700 \\ 
J103628.12+501157.9 & 2.0097 & MgII & J103630.52+501219.8 & 2.228 & BOSS & 271 & 1198 \\ 
J103857.37+502707.0 & 3.1325 & [OIII] & J103900.01+502652.8 & 3.236 & ESI & 233 & 3567 \\ 
J103946.92+454716.0 & 1.8644 & MgII & J103945.58+454707.4 & 2.456 & BOSS & 148 & 675 \\ 
J104244.84+650002.7 & 1.9876 & MgII & J104245.14+645936.7 & 2.124 & BOSS & 227 & 703 \\ 
J104435.62+313950.7 & 1.7062 & MgII & J104434.76+313957.7 & 2.377 & BOSS & 115 & 1873 \\ 
J104955.01+231358.2 & 1.8439 & MgII & J104953.97+231401.3 & 2.171 & BOSS & 129 & 1813 \\ 
J105221.77+555253.5 & 1.9989 & MgII & J105218.36+555311.3 & 2.278 & BOSS & 293 & 846 \\ 
J105246.45+641832.2 & 1.6429 & MgII & J105251.42+641838.5 & 2.936 & BOSS & 286 & 127 \\ 
J111339.86+330604.8 & 1.8913 & MgII & J111337.84+330553.3 & 2.413 & BOSS & 243 & 853 \\ 
J111850.44+402553.8 & 1.9257 & MgII & J111851.45+402557.6 & 2.317 & BOSS & 106 & 1946 \\ 
J112858.89+644440.4 & 1.6561 & MgII & J112854.14+644427.4 & 2.217 & BOSS & 289 & 149 \\ 
J113852.65+632934.0 & 1.8855 & MgII & J113851.73+632955.6 & 2.625 & BOSS & 196 & 1912 \\ 
J114435.54+095921.7 & 2.9734 & [OIII] & J114436.65+095904.9 & 3.16 & MIKE-Red & 189 & 2914 \\ 
J114439.51+454115.8 & 1.687 & MgII & J114442.48+454111.3 & 2.592 & BOSS & 275 & 78 \\ 
J114546.54+032236.7 & 1.7664 & MgII & J114546.22+032251.9 & 2.011 & MagE & 139 & 779 \\ 
J115253.09+150706.5 & 1.7883 & MgII & J115254.97+150707.8 & 3.349 & BOSS & 237 & 622 \\ 
J115457.16+471149.3 & 1.6819 & MgII & J115458.69+471209.9 & 1.947 & SDSS & 226 & 60 \\ 
J115502.45+213235.5 & 1.9551 & MgII & J115504.25+213254.0 & 2.695 & BOSS & 277 & 397 \\ 
J115529.49+463413.1 & 1.6491 & MgII & J115528.75+463442.9 & 2.329 & BOSS & 270 & 326 \\ 
J115533.62+393359.2 & 1.6118 & MgII & J115531.32+393415.4 & 2.555 & BOSS & 272 & 742 \\ 
J120224.68+074800.3 & 1.6613 & MgII & J120226.48+074739.7 & 2.767 & BOSS & 296 & 584 \\ 
J120417.47+022104.7 & 2.436 & [OIII] & J120416.69+022110.0 & 2.532 & HIRES & 112 & 2710 \\ 
J120856.94+073741.2 & 2.1708 & MgII & J120857.16+073727.3 & 2.616 & MagE & 123 & 3853 \\ 
J121159.88+324009.0 & 1.978 & MgII & J121201.69+324013.3 & 2.273 & BOSS & 209 & 1046 \\ 
J121344.28+471958.7 & 1.8371 & MgII & J121343.01+471931.0 & 3.275 & BOSS & 260 & 491 \\ 
J1215590+571616.6 & 1.93 & [OIII] & J121558.82+571555.5 & 1.964 & BOSS & 184 & 614 \\ 
J121657.82+152706.6 & 1.9473 & MgII & J121657.00+152712.7 & 2.318 & BOSS & 116 & 4225 \\ 
J122514.29+570942.3 & 1.8953 & MgII & J122517.89+570943.7 & 2.224 & BOSS & 255 & 330 \\ 
J123143.01+002846.3 & 3.2015 & [OIII] & J123141.73+002913.9 & 3.308 & GMOS-S & 271 & 1490 \\ 
J124632.33+234531.2 & 1.9937 & MgII & J124632.19+234509.5 & 2.573 & BOSS & 188 & 1886 \\ 
J124846.05+405758.2 & 1.8265 & MgII & J124846.97+405820.9 & 2.463 & BOSS & 219 & 897 \\ 
J130124.74+475909.6 & 2.194 & H$\alpha$ & J130125.67+475930.8 & 2.765 & SDSS & 199 & 4932 \\ 
J130605.19+615823.7 & 2.1089 & H$\alpha$ & J130603.55+615835.2 & 2.175 & LRIS & 141 & 1761 \\ 
J130714.79+463536.6 & 1.6226 & MgII & J130716.07+463511.2 & 2.248 & BOSS & 251 & 137 \\ 
J131341.32+454654.6 & 1.6878 & MgII & J131342.78+454658.2 & 2.241 & BOSS & 139 & 1098 \\ 
J132514.97+540930.6 & 2.0507 & MgII & J132511.07+540927.0 & 3.235 & BOSS & 298 & 514 \\ 
J133026.12+411432.0 & 2.0645 & MgII & J133023.67+411445.9 & 2.217 & BOSS & 271 & 384 \\ 
J133924.02+462808.2 & 1.8539 & MgII & J133922.31+462749.2 & 3.391 & BOSS & 226 & 940 \\ 
J134650.08+195235.2 & 2.0697 & MgII & J134648.19+195253.1 & 2.523 & BOSS & 278 & 884 \\ 
J135306.35+113804.7 & 1.6315 & MgII & J135307.90+113805.5 & 2.431 & BOSS & 213 & 8963 \\ 
J135849.71+273806.9 & 1.9008 & MgII & J135849.54+273756.9 & 2.127 & LRIS & 89 & 1765 \\ 
J140208.01+470111.1 & 1.9161 & [OIII] & J140209.52+470117.8 & 2.269 & BOSS & 140 & 1437 \\ 
J140918.01+522552.4 & 1.8808 & MgII & J140916.98+522535.3 & 2.109 & SDSS & 170 & 583 \\ 
J141337.18+271517.1 & 1.6905 & MgII & J141337.96+271511.0 & 1.965 & BOSS & 105 & 609 \\ 
J142003.67+022726.7 & 3.617 & [OIII] & J142004.12+022708.8 & 4.191 & ESI & 144 & 2291 \\ 
J142054.42+160333.3 & 2.0221 & [OIII] & J142054.92+160342.9 & 2.057 & MagE & 104 & 7811 \\ 
J142215.57+465230.7 & 1.748 & MgII & J142214.63+465254.6 & 2.338 & BOSS & 225 & 639 \\ 
J142758.89$-$012130.4 & 2.2738 & [OIII] & J142758.74-012136.2 & 2.354 & MIKE & 53 & 34208 \\ 
J143109.67+572728.0 & 1.6802 & MgII & J143109.22+572726.4 & 2.063 & BOSS & 39 & 5166 \\ 
J143312.56+082651.8 & 1.8807 & MgII & J143313.99+082714.0 & 2.432 & BOSS & 274 & 531 \\ 
J143345.55+064109.0 & 2.294 & [OIII] & J143344.55+064111.9 & 2.34 & BOSS & 122 & 1130 \\ 
J143609.15+313426.7 & 1.8774 & MgII & J143610.68+313418.9 & 2.562 & BOSS & 183 & 2188 \\ 
J144211.25+530252.0 & 1.6461 & MgII & J144209.98+530308.0 & 2.632 & BOSS & 179 & 439 \\ 
J144232.92+013730.4 & 1.8079 & MgII & J144231.91+013734.8 & 2.274 & BOSS & 137 & 1988 \\ 
J144429.34+311321.2 & 1.7355 & MgII & J144427.96+311313.0 & 1.795 & LRIS & 173 & 8820 \\ 
J150814.06+363529.4 & 1.8493 & MgII & J150812.78+363530.3 & 2.105 & BOSS & 133 & 5111 \\ 
J153328.83+142542.5 & 2.0782 & MgII & J153329.17+142537.8 & 2.564 & BOSS & 59 & 3898 \\ 
J153456.02+215342.3 & 1.6712 & MgII & J153455.85+215324.7 & 2.529 & BOSS & 156 & 1702 \\ 
J153954.74+314629.3 & 1.8747 & MgII & J153952.46+314625.2 & 2.235 & BOSS & 256 & 1080 \\ 
J155325.61+192140.0 & 2.01 & [OIII] & J155325.89+192137.7 & 2.098 & MagE & 44 & 5576 \\ 
J155422.88+124438.0 & 1.8169 & MgII & J155424.39+124431.5 & 2.394 & BOSS & 202 & 1349 \\ 
J155947.73+494307.3 & 1.8615 & MgII & J155946.28+494326.7 & 1.945 & LRIS & 210 & 1743 \\ 
J160547.61+511330.5 & 1.783 & MgII & J160546.67+511322.9 & 1.844 & LRIS & 102 & 4045 \\ 
J161930.94+192620.9 & 1.7821 & MgII & J161929.78+192645.4 & 2.39 & BOSS & 258 & 538 \\ 
J162738.63+460538.4 & 3.8149 & [OIII] & J162737.25+460609.3 & 4.11 & ESI & 253 & 1959 \\ 
J163121.74+433317.3 & 2.0182 & MgII & J163123.57+433317.3 & 2.631 & BOSS & 172 & 4090 \\ 
J165442.21+251249.2 & 1.7207 & MgII & J165444.38+251306.2 & 2.341 & BOSS & 298 & 459 \\ 
J165716.85+310513.0 & 2.1331 & MgII & J165716.52+310524.5 & 2.395 & MODS1 & 98 & 5222 \\ 
J214620.69$-$075250.6 & 2.1155 & [OIII] & J214620.99-075303.8 & 2.577 & MagE & 120 & 4869 \\ 
J214813.26+263059.4 & 1.6285 & MgII & J214814.36+263129.7 & 3.286 & BOSS & 295 & 78 \\ 
J220248.61+123645.5 & 2.0697 & [OIII] & J220248.31+123656.3 & 2.512 & BOSS & 100 & 9466 \\ 
J233845.19$-$000327.1 & 2.4399 & [OIII] & J233845.45-000331.8 & 2.997 & XSHOOTER & 51 & 3674 \\ 
J235505.22+320058.0 & 1.8159 & MgII & J235505.33+320105.4 & 2.367 & BOSS & 56 & 2271 \\ 
J235819.92+342455.8 & 1.6235 & MgII & J235819.33+342506.5 & 2.02 & BOSS & 113 & 480 \\ 
\enddata 
\tablenotetext{a}{The emission-line analyzed for measuring $z_{\rm fg}$.} 
\end{deluxetable*}

\begin{figure}
\includegraphics[width=3in]{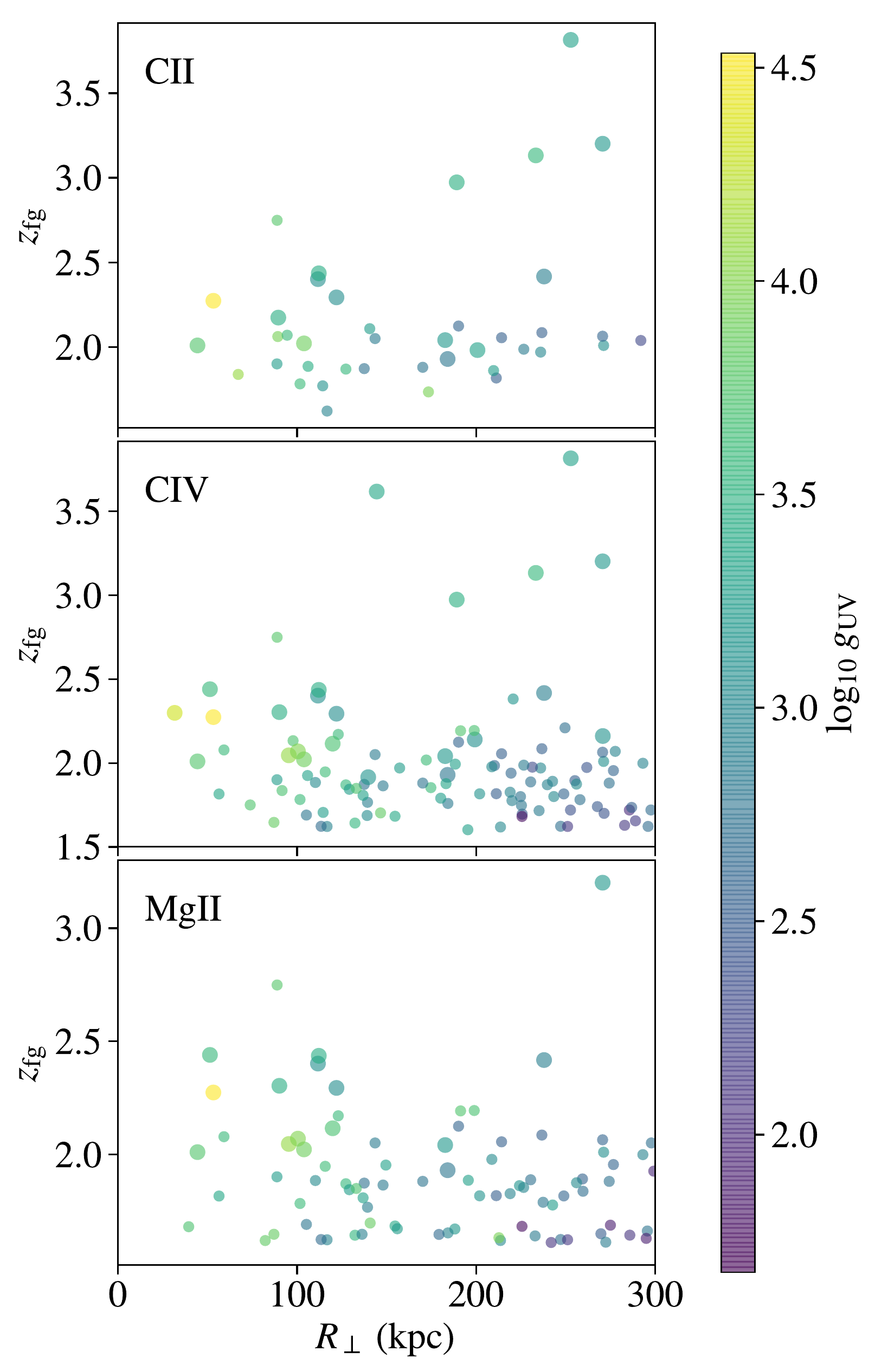}
\caption{These panels summarize properties of the QPQ9 dataset. The QPQ survey selects
quasar pairs of $R_\perp < 300$\,kpc and $z_{\rm fg}>1.6$. Assuming that the foreground quasars
emit isotropically and at a distance equal to the impact parameter, the enhancement in the UV
flux relative to the extragalactic UV background, $g_{\rm UV}$, can be estimated. Large
symbols correspond to foreground quasars with the most precise redshift measurement from
[\ion{O}{3}]\,5007, while small symbols correspond to $z_{\rm fg}$
measurements from \ion{Mg}{2}\,2800, H$\alpha$, or H$\beta$ emission. The top panel shows
quasar pairs with coverage of \ion{C}{2}\,1334 at $z_{\rm fg}$ in the background quasar spectra.
The middle panel shows pairs with coverage of \ion{C}{4}\,1548. The bottom panel shows
pairs with coverage of \ion{Mg}{2}\,2796.
}
\label{fig:experiment}
\end{figure}

The goal of our experiment is to measure the average velocity fields of the absorption from C$^+$,
C$^{3+}$, and Mg$^+$ ions associated with the CGM of the galaxies hosting $z\sim 2$ quasars.

From our QPQ survey\footnote{http://www.qpqsurvey.org}, we analyze a subset of systems that
pass within transverse separation $R_\perp<300$\,kpc from a foreground quasar with
$z_{\rm fg} > 1.6$. We restrict the sample to foreground quasars with redshift measured from
\ion{Mg}{2}\,2800, [\ion{O}{3}]\,5007, or H$\alpha$ emission, giving a precision of
$300\,{\rm km\,s^{-1}}$ or better and an average offset from the systemic redshift of
$|100\,{\rm km\,s^{-1}}|$ or less. According to \cite{Shen+16}, the [\ion{O}{3}] emission-line
redshifts have the smallest scatter (intrinsic scatter and measurement error combined) of
$68\,{\rm km\,s^{-1}}$ about the systemic redshift, and we analyze the sub-sample with
[\ion{O}{3}] redshifts separately. The [\ion{O}{3}] line has an average blueshift of
$48\,{\rm km\,s^{-1}}$ about the systemic redshift, which has been added when we compute the
redshift of the line. The scatter and average offset of [\ion{O}{3}] redshifts reported by
\cite{Shen+16} is consistent with the numbers reported by \cite{Boroson05} using a larger but
lower redshift sample. Systemic redshifts measured from \ion{Mg}{2} have a precision of
$226\,{\rm km\,s^{-1}}$ according to \cite{Shen+16}, and we have taken into account their
reported median blueshift of $57\,{\rm km\,s^{-1}}$ of \ion{Mg}{2} from the systemic. We note that
\cite{Richards+02} reported a median redshift of $97\,{\rm km\,s^{-1}}$ of \ion{Mg}{2} from
[\ion{O}{3}] using a larger but lower redshift sample. In QPQ8, we quantified
the precision of H$\alpha$ to be $300\,{\rm km\,s^{-1}}$ and the median offset from the systemic
redshift is close to zero, consistent with the velocity shifts measured by \cite{Shen+11}.
Although H$\beta$ is a narrow emission-line, we do not consider its redshift sufficiently reliable
for use as systemic redshift. H$\beta$ redshifts have a large scatter about the systemic
$\approx400\,{\rm km\,s^{-1}}$, and a large average offset about the systemic
$\approx100\,{\rm km\,s^{-1}}$ \citep[][QPQ8]{Shen+16}.
Our line-centering algorithm calculates the mode of a line given by
$3\times{\rm median}-2\times{\rm mean}$, applied to the upper 60\% of the emission, while
\cite{Shen+16} calculates the peak of a line. We expect that our line-centering algorithm gives
emission redshifts very comparable to the \cite{Shen+16} algorithm, however. \cite{Shen+16}
states that the difference between the peak and the centroid of an emission-line
is not significant except for the broad line H$\beta$, which we do not use in redshift
measurements. To quantify the above, we further obtain individual measurements of centroids and
peaks in the \cite{Shen+16} sample through private communication. We found there is essentially no
difference between using the centroid versus using the peak for [\ion{O}{3}] emission redshifts,
and there is on average $50\,{\rm km\,s^{-1}}$ difference for \ion{Mg}{2}. We may expect the
difference between the mode and the peak is even smaller. Hence, we argue that the average
systemic bias corrections measured in \cite{Shen+16} may be self-consistently applied to our
measured emission-line redshifts to obtain systemic redshifts.

We further add to the QPQ dataset with quasar pairs selected from the public dataset of
igmspec\footnote{https://github.com/specdb/igmspec} \citep{Prochaska+17}, which includes the
spectra from the quasar catalogs based upon the Sloan Digital Sky Survey Seventh Data Release
\citep{Schneider+10} and the Twelfth Data Release \citep{Paris+17}. We only select pairs with
$z_{\rm fg}$ measurable using a robust \ion{Mg}{2}\,2800 emission-line.
We reach a final sample size of 148.
Figure~\ref{fig:experiment} summarizes the experimental design. We refer the reader to previous
QPQ publications for the details on the emission-line centering algorithm, data reduction, and
continuum normalization (QPQ1; QPQ6; QPQ8).

As in the previous QPQ papers, we select the quasar pairs to have redshift difference
$>3000\,{\rm km\,s^{-1}}$, to exclude physically associated binary quasars. The cut on velocity
difference is motivated by the typical redshift uncertainty of $\approx500\,{\rm km\,s^{-1}}$ of
the background quasars. In QPQ8, it was required that the observed wavelengths of the
metal ion transitions fall outside the Ly$\alpha$ forest of the background quasar. In this paper,
we exclude a small window around the Ly$\alpha$ emission, in additional to the Ly$\alpha$ forest,
from analysis. For stacked profile analysis, a good estimate of
the continuum level is necessary. In QPQ8 we found that absorption associated to
the foreground quasar occurs within $\pm2000\,{\rm km\,s^{-1}}$ around $z_{\rm fg}$. Therefore, it
is desirable to keep a $\approx\pm3000\,{\rm km\,s^{-1}}$ window relatively free of contamination
from Ly$\alpha$ forest. Taking into account the redshift uncertainties,
we decide that at least one transition among \ion{C}{2}\,1334, \ion{C}{4}\,1548, and
\ion{Mg}{2}\,2796 at $z_{\rm fg}$ must lie redward of
$(1215.6701+20)\times(1+z_{\rm bg})\,{\rm \AA}$, for a pair to be included in the analysis.

Furthermore, we include only those spectra with average signal-to-noise ratio (S/N) exceeding 5.5
per rest-frame \AA\ in a $\pm3000\,{\rm km\,s^{-1}}$ window centered on the observed wavelengths
of the metal ion transitions. This criterion is a compromise between maximizing sample size versus
maintaining good data quality on the individual sightlines. We find that S/N $>5.5$ per rest-frame
\AA\ is necessary for properly estimating the continuum, as well as identifying mini-broad
absorption line systems associated to the background quasar, which will significantly depress the
flux level. We also require that the region of the spectrum that is $\pm3000\,{\rm km\,s^{-1}}$
around a considered metal ion transition does not overlap with strong atmospheric ${\rm O}_2$
bands. The ${\rm O}_2$ A- and B-band span 7595\textendash7680\,\AA \ and 6868\textendash6926\,\AA
\ respectively.

Table~\ref{tab:sample} lists the full QPQ9 sample. In Table~\ref{tab:summary}, we first list the
sample size, the median $z_{\rm fg}$, and the median $R_\perp$ of the quasar pairs that survive
the above selection criteria for \ion{C}{2}\,1334, \ion{C}{4}\,1548, and \ion{Mg}{2}\,2796
respectively. We then provide the summary for the sub-sample with $z_{\rm fg}$ measured from
[\ion{O}{3}].

\section{Analysis}
\label{sec:analysis}

\subsection{Stacked Profiles}
\label{sec:stacks}

\begin{figure}
\includegraphics[width=3.5in]{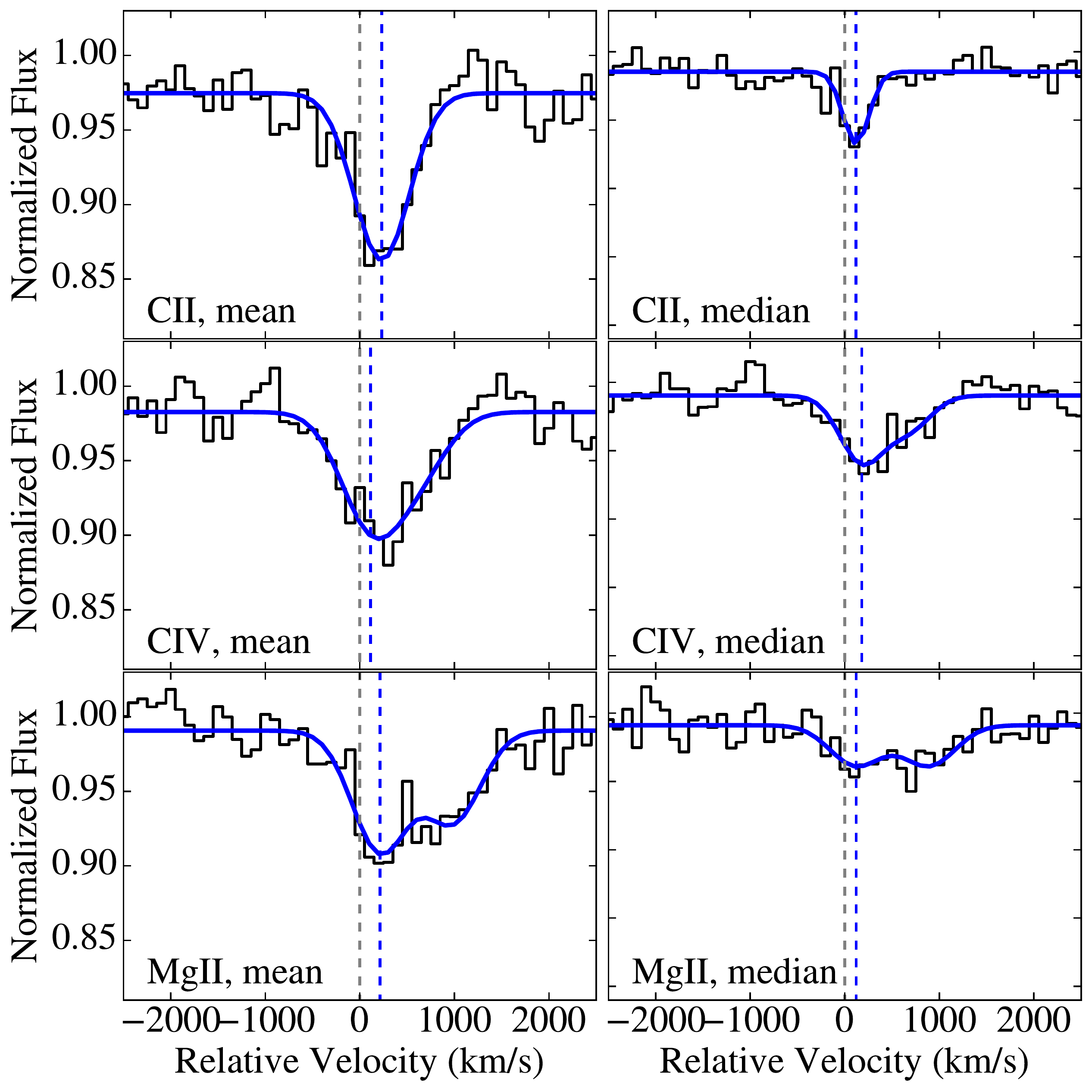}
\caption{Mean and median absorption centered at \ion{C}{2}\,1334, \ion{C}{4}\,1548, and
\ion{Mg}{2}\,2796 of the foreground quasars for all QPQ9 pairs. The composites are shown in thin,
black. Overplotted on the composites are Gaussian fits, normalized to pseudo-continua far away
from a velocity of $0\,{\rm km\,s^{-1}}$ relative to $z_{\rm fg}$, and are shown in thick, blue.
For the doublets, a second Gaussian with a fixed mean separation and a tied standard deviation is
included in the modeling. The absorptions frequently exhibit large velocity widths. The blue dashed
lines mark the centroids, which show small positive velocity offsets from $z_{\rm fg}$. The
$1\sigma$ modeling error for the
dispersion and the centroid of the \ion{C}{2} mean stack are $27\,{\rm km\,s^{-1}}$ and
$25\,{\rm km\,s^{-1}}$ respectively, which are}
also the typical modeling errors of the other stacks.
\label{fig:stacks_and_fits}
\end{figure}

We create composite spectra that average over the intrinsic scatter in quasar environments,
continuum placement errors, and redshift errors. The individual spectra of background quasars are
shifted to the rest-frame of the foreground quasars at the transitions of interest. Each spectrum
has been linearly interpolated onto a fixed velocity grid centered at $z_{\rm fg}$ with bins of
$100\,{\rm km\,s^{-1}}$. For a velocity bin of this size, it is unnecessary to smooth the data to
a common spectral resolution.
The individual spectra are then combined with a mean or a median
statistic.
Spectra with broad absorption-line systems and mini-broad absorption-line systems are excluded.
Bad pixels in the individual spectra have been masked before generating the
composites. Since each quasar pair gives an independent probe of the CGM, each pair has an equal
weighting in the stacked profiles, i.e.\ we do not weight the spectra by the measured S/N near the
metal ion transitions. Scatter in the stacked spectra is dominated by randomness in the CGM rather
than scatter in the flux of individual observations. The mean statistic of the individual spectra
yields a good estimate of the average absorption and preserves equivalent width. The median
statistic is less sensitive to outliers. However, the median opacity at any velocity channel is
rather small, since the discrete absorbers are spread throughout the entire velocity window.
A pixel is not affected unless there are more than 50\% of quasar pairs with absorption at its
velocity. The analysis on the median velocity field is thus subject to larger uncertainty. In the
following we present stacked spectra using both the mean and the median statistic.

In Figure~\ref{fig:stacks_and_fits}, we present mean and median stacks of \ion{C}{2}\,1334,
\ion{C}{4}\,1548, and \ion{Mg}{2}\,2796 absorption of the QPQ9 sample. We focus on the
analysis results of the \ion{C}{2} mean stack. \ion{C}{4} and \ion{Mg}{2} are doublet transitions
and it is more challenging to analyze their kinematics.
Two results are evident in Figure~\ref{fig:stacks_and_fits}: (i) the mean \ion{C}{2} stack
exhibits excess absorption spanning a large velocity width; (ii) the mean absorption is likely
skewed toward positive velocities.

Visually, there are two absorption components. One component is the uniform depression in the
continuum level in the stack resulting from absorbers unassociated with the foreground
quasars. The other component comes from absorbers associated with the foreground quasars and
distribute around their systemic velocities. To model the absorption, we introduce
a Gaussian profile while allowing a constant ``pseudo-continuum'' to vary.
We perform $\chi^2$ minimization with each channel given equal weight. From the best-fit
to the data, we measure
the $1\sigma$ dispersion of the stack to be $293\,{\rm km\,s^{-1}}$ and the centroid of the
\ion{C}{2} stack to be $+232\,{\rm km\,s^{-1}}$. The dispersion suggests extreme kinematics, while
the centroid suggests an asymmetry that contradicts the standard expectation.
The median
stack, on the other hand, shows weaker absorption,
and the Gaussian model has a more uncertain dispersion and a centroid with smaller offset.

To test whether the dispersion in the average absorption associated with the foreground quasars is
well-captured by a Gaussian model, we also calculate the dispersion separately using the standard
deviation formula. Specifically, we apply the standard deviation formula to a $\pm1300\,{\rm km\,s^{-1}}$
window surrounding the absorption centroid, while fitting a continuum to the rest of the
$\pm3000\,{\rm km\,s^{-1}}$ window for stacking. The
dispersion measured with the above method is essentially the same as that found by fitting a Gaussian.
One may also speculate on the existence of a broader absorption component hidden in the depressed
continuum. We try replacing the constant absorption component with a broad Gaussian component in our
modeling. We find that this second Gaussian component has
an amplitude of $0.030\pm0.009$, i.e.\ essentially the same amplitude as the initial
constant absorption component, and a $1\sigma$ dispersion of $3614\pm1731\,{\rm km\,s^{-1}}$, i.e.\ wider
than the entire velocity window for stacking.
This weak and very broad component has no effect on the width and the centroid of the narrow,
associated Gaussian component. We therefore consider that one single Gaussian component is a good
description of the average absorption associated with the foreground quasars.
Moreover, narrow associated absorbers of background quasars will not prefer the systemic velocities of
the foreground quasars, and hence will not affect the average absorption measured for the foreground
quasars. Their contribution to the average absorption should result in a tilt in the
pseudo-continuum, which is insignificant.

We also create mean and median stack for the sub-sample with [\ion{O}{3}] redshifts and model the
absorption with Gaussian best-fit. The \ion{C}{2} mean stack for this sub-sample has
a dispersion of $330\,{\rm km\,s^{-1}}$, and a centroid at $+235\,{\rm km\,s^{-1}}$,
consistent with the full sample.

To model the mean and median absorption of \ion{C}{4}\,1548 and \ion{Mg}{2}\,1796, we introduce a
second Gaussian with separation equal to the doublet separation ($498\,{\rm km\,s^{-1}}$ and
$769\,{\rm km\,s^{-1}}$ respectively), and tie the dispersion of the two lines in a doublet. We
allow the doublet ratio to vary from 2:1 to 1:1.
The centroid is degenerate with the doublet ratio in our modeling. This degeneracy is more
pronounced for \ion{C}{4}, although we find that a double ratio closer to 2:1 better captures the
absorption at the central few pixels. We state that the modeling of the doublets is presented for
consistency check, but the analysis focuses on \ion{C}{2}.
The modeling results show that the velocity fields
of \ion{C}{4} and \ion{Mg}{2} are consistent with \ion{C}{2}, i.e.\ large dispersion and centroid
is skewed toward positive velocities.

The above analyses are summarized in Table~\ref{tab:summary}. The Gaussian models normalized to
pseudo-continuum are overplotted on the data stacks in Figure~\ref{fig:stacks_and_fits}.

\begin{deluxetable*}{lccc} 
\tablewidth{0pc} 
\tablecaption{Summary of the Data and Analysis\label{tab:summary}} 
\tabletypesize{\small} 
\tablehead{\colhead{Measure} & \colhead{\ion{C}{2}\,1334} 
& \colhead{\ion{C}{4}\,1548} & \colhead{\ion{Mg}{2}\,2796}} 
\startdata 
\cutinhead{For the Full QPQ9 Sample} 
Number of pairs & 40 & 110 & 86 \\ 
Median $z_{\rm fg}$ & 2.04 & 1.89 & 1.87 \\ 
Median $R_\perp$ & 157 & 191 & 184 \\ 
1$\sigma$ dispersion of mean stack (${\rm km\,s^{-1}}$) & $293\pm87$ & $303\pm44$ & $295\pm200$ \\ 
Centroid of mean stack (${\rm km\,s^{-1}}$) & $+232\pm98$ & $+11\pm63$ & $+215\pm124$ \\ 
Pseudo-continuum of mean stack & $0.97$ & $0.98$ & $0.99$ \\ 
1$\sigma$ dispersion of median stack (${\rm km\,s^{-1}}$) & $137\pm569$ & $218\pm71$ & $276\pm215$ \\ 
Centroid of median stack (${\rm km\,s^{-1}}$) & $+119\pm240$ & $+189\pm81$ & $+121\pm201$ \\ 
Pseudo-continuum of median stack & $0.99$ & $0.99$ & $0.99$ \\ 
\cutinhead{For the Sub-sample with [OIII] Redshifts} 
Number of pairs & 15 & 23 & 15 \\ 
Median $z_{\rm fg}$ & 2.29 & 2.29 & 2.27 \\ 
Median $R_\perp$ & 183 & 122 & 112 \\ 
1$\sigma$ dispersion of mean stack (${\rm km\,s^{-1}}$) & $330\pm103$ & $255\pm41$ & $235\pm129$ \\ 
Centroid of mean stack (${\rm km\,s^{-1}}$) & $+235\pm118$ & $+130\pm77$ & $+210\pm119$ \\ 
Pseudo-continuum of mean stack & $0.98$ & $0.98$ & $0.95$ \\ 
1$\sigma$ dispersion of median stack (${\rm km\,s^{-1}}$) & $103\pm236$ & $211\pm71$ & $175\pm114$ \\ 
Centroid of median stack (${\rm km\,s^{-1}}$) & $+256\pm191$ & $+210\pm111$ & $+244\pm125$ \\ 
Pseudo-continuum of median stack & $0.99$ & $0.98$ & $0.98$ \\ 
\enddata 
\end{deluxetable*}

\subsection{Interpretation of the large velocity fields}
\label{sec:significance_width}

\begin{figure}
\includegraphics[width=2.5in]{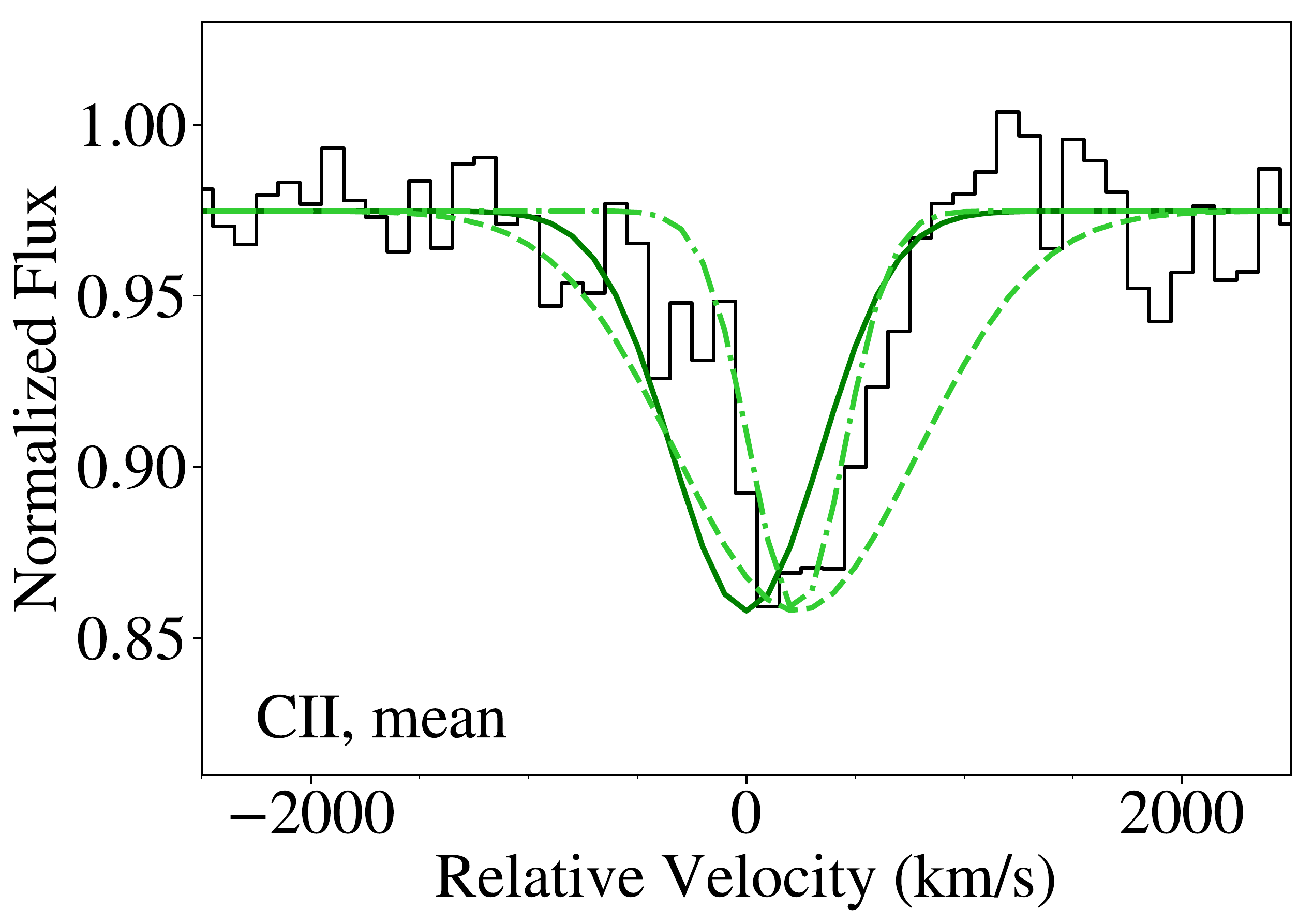}
\caption{The same \ion{C}{2} mean stack shown in the first panel of
Figure~\ref{fig:stacks_and_fits} is shown in thin, black.
We overplot in dashed, limegreen an
absorption profile with $\sigma_v=554\,{\rm km\,s^{-1}}$, which is larger than the observed width
by three times the standard deviation in the bootstrap analysis. Motions that produce a velocity
width larger than this can be ruled out. We overplot in dot dashed, limegreen an absorption profile with
$\sigma_v=214\,{\rm km\,s^{-1}}$, which is smaller than the observed width by three times the
modeling error. Unless gravitational and Hubble flows together with redshift error broadening
produce a velocity width smaller than this, extra dynamical processes (e.g.\ outflows) will not be
required to explain the observed width.
In thick, green, we overplot the
Gaussian absorption model of the Monte Carlo simulations generated from a purely clustering
argument. The model from clustering analysis is multiplied to the pseudo-continuum level of the
stack of the observational data, and broadened
by the mean redshift error in the data. While this model has a dispersion within modeling error of the
dispersion in the data. the centroid of the stack of the data appears to be redshifted
from the systemic. A model with only gravitational motions and Hubble flows cannot explain
this putative asymmetry.
}
\label{fig:monte}
\end{figure}

\begin{figure}
\includegraphics[width=2.5in]{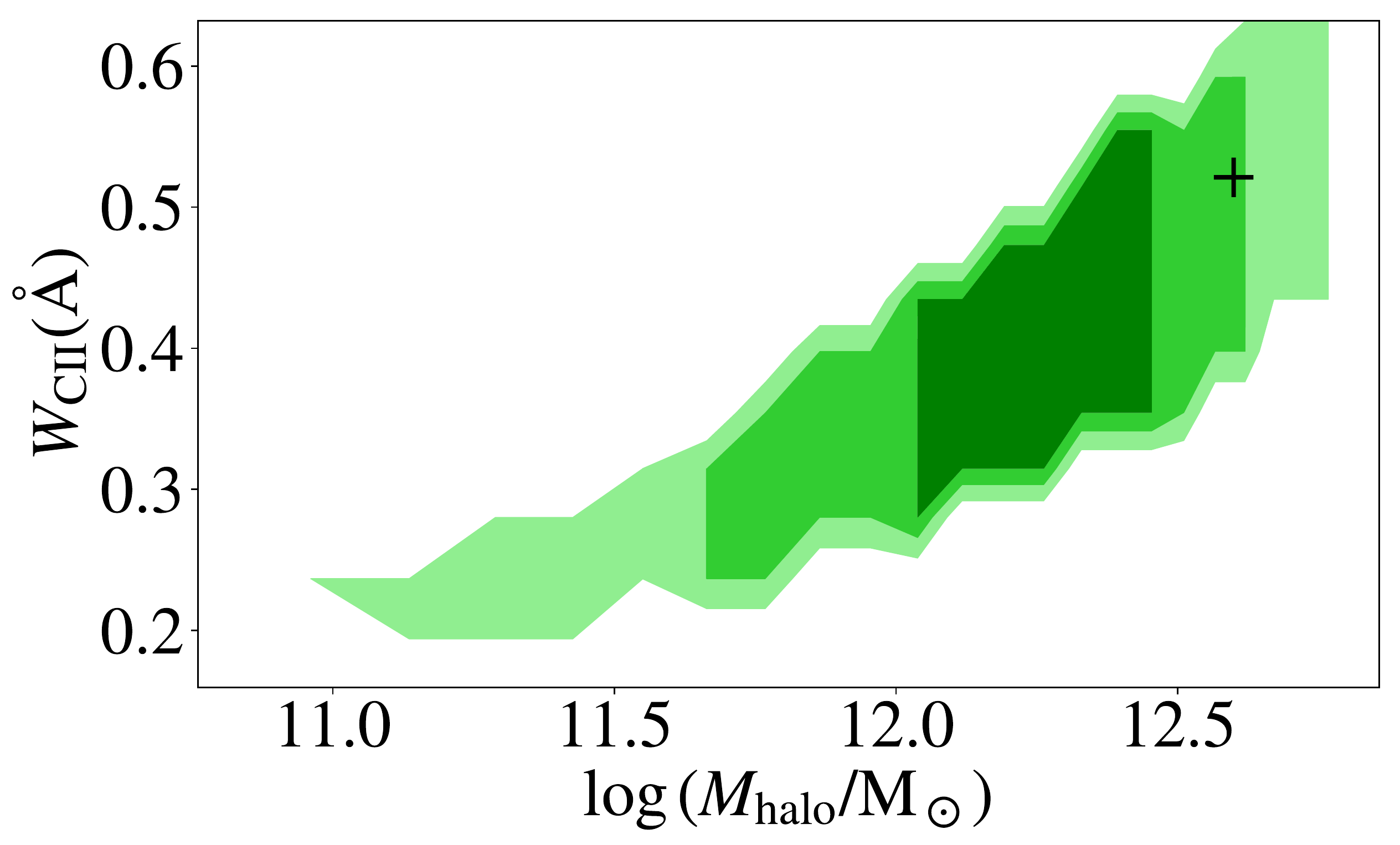}
\caption{
Probability distributions of the parameters $W_{\rm CII}$ and $M_{\rm halo}$. The plot shows the
degeneracy between $W_{\rm CII}$ and $M_{\rm halo}$ in recovering the intrinsic width of the
absorption profile. We mark contours for points that produce an absorption
profile of width that is 1, 2, and 3 times the modeling error away from the observed intrinsic
width. The intrinsic velocity width corresponding to typical QPQ halo mass is marked with a plus
sign, and is contained within the 2$\sigma$ contour.}
\label{fig:contour}
\end{figure}

Under the assumption that the intrinsic dispersion and the redshift uncertainty add in
quadrature to give the observed width, we solve for the intrinsic dispersion in the \ion{C}{2}
mean stack. For the full QPQ9 sample, with the mean
$\sigma_{{\rm error}(z)}^{\rm full}=189\,{\rm km\,s^{-1}}$, we recover
$\sigma_{\rm intrinsic}^{\rm full}=224\,{\rm km\,s^{-1}}$. For the sub-sample with [\ion{O}{3}]
redshifts, we recover $\sigma_{\rm intrinsic}^{\rm [OIII]}=323\,{\rm km\,s^{-1}}$.

To assess the statistical significance of the observed dispersion, we perform a bootstrap
analysis by randomly resampling from the full sample 10000 times. We introduce a Gaussian absorption
profile to model each bootstrap realization. We find that a width of
$\sigma_v>554\,{\rm km\,s^{-1}}$ would be larger than the observed by three times the scatter in
the bootstrap realizations
(overplotted in Figure~\ref{fig:monte}). In other words, taking into account the redshift errors,
motions in addition to gravitational and Hubble flows that will produce an intrinsic dispersion
$\sigma_v>521\,{\rm km\,s^{-1}}$ can be ruled out.

On the contrary,
$\sigma_{\rm rms}<70\,{\rm km\,s^{-1}}$ together with Hubble velocities and
broadening by redshift errors will result in a velocity width that is more than three times the
modeling error away from the observed width (overplotted in Figure~\ref{fig:monte}). This implies
that, unless the characteristic $M_{\rm halo}<10^{11.0}\,M_\odot$,
additional dynamical processes are not required to explain the observed width.

\cite{Eftekharzadeh+15} measured the clustering of quasars in the range $2.2<z<2.8$ while
\cite{RodriguezTorres+17} measured the clustering of quasars in the range $1.8<z<2.2$. They
estimated that these quasars are hosted by dark matter halos with mass
$M_{\rm halo}=10^{12.5}\,M_\odot$ and $M_{\rm halo}=10^{12.6}\,M_\odot$ respectively. If dark
matter halos hosting QPQ9 quasars have a characteristic mass $10^{12.6}\,M_\odot$ and follow an
NFW profile \citep{NavarroFrenkWhite97} with concentration parameter $c=4$, at $z\approx2$ the
maximum circular velocity is $345\,{\rm km\,s^{-1}}$. \cite{TormenBouchetWhite97} found that the
maximum circular velocity is $\approx1.4$ times the maximum of the average one-dimensional
root-mean-square velocity $\sigma_{\rm rms}$. Hence, the average line-of-sight rms velocity
typical of QPQ9 halos is $\sigma_{\rm rms}=246\,{\rm km\,s^{-1}}$. In QPQ8, we estimated the
probability of intercepting a random optically thick absorber is 4\%, and clustering would only
increase that to 24\%. Although motions due to Hubble flows do not dominate, they nevertheless
contribute to the observed dispersion. We can investigate
whether gravitational motions and Hubble flows are sufficient to reproduce the dispersion in the
data, using Monte Carlo methods to simulate the absorption signals.

Since \ion{C}{2} systems arise in optically thick absorbers, we may adopt the clustering analysis
results of QPQ6. In the absence of clustering, the expected number of absorbers per unit redshift
interval for Lyman limit systems, super Lyman limit systems, and damped Ly$\alpha$ systems are
respectively $\ell_{\rm IGM}^{\rm LLS}(z)\approx1.05((1+z)/(1+2.5))^{2.1}$,
$\ell_{\rm IGM}^{\rm SLLS}(z)\approx0.44((1+z)/(1+2.5))^{2.1}$, and
$\ell_{\rm IGM}^{\rm DLA}(z)\approx0.2((1+z)/(1+2.5))^{2.1}$. The quasar-absorber correlation
functions for Lyman limit systems, super Lyman limit systems, and damped Ly$\alpha$ systems are
respectively $\xi_{\rm QA}^{\rm LLS}(r)=(r/(12.5\,h^{-1}\,{\rm Mpc}))^{-1.68}$,
$\xi_{\rm QA}^{\rm SLLS}(r)=(r/(14.0\,h^{-1}\,{\rm Mpc}))^{-1.68}$, and
$\xi_{\rm QA}^{\rm DLA}(r)=(r/(3.9\,h^{-1}\,{\rm Mpc}))^{-1.6}$.
For each quasar pair, we calculate the expected number of optically thick absorbers within
$\pm3000\,{\rm km\,s^{-1}}$ at a distance $R_\perp$ from the foreground quasar and at
$z_{\rm fg}$. Then we generate 1000 mock sightlines. The number of absorbers for each mock
spectrum is randomly selected from a Poisson distribution with mean equal to the expected number
calculated as above. The absorbers are randomly assigned Hubble velocities, with a probability
distribution according to the quasar-absorber correlation functions. The absorbers are randomly
assigned additional peculiar velocities drawn from a normal distribution with mean equal to
$0\,{\rm km\,s^{-1}}$ and scatter equal to $\sigma_{\rm rms}$. For each absorber, we assume a rest
equivalent width for \ion{C}{2} $W_{\rm CII}$ and a Gaussian absorption profile. We repeat the
above procedure for all 40 quasar pairs, and create a mean stack of the 40000 mock spectra
generated. We fit a Gaussian absorption profile multiplied to a constant continuum level to model
the stack of mock spectra. We adjust the $W_{\rm CII}$ adopted for the absorbers until the
amplitude of the best-fit Gaussian of the stack of mock spectra matches the amplitude of the stack
of the observational data. We find that $W_{\rm CII}\approx0.5\,{\rm \AA}$ well reproduces the
amplitude, and the dispersion of the Gaussian absorption model is insensitive to the assumed
$W_{\rm CII}$ or line profile for one absorber.

In Figure~\ref{fig:monte}, we show a comparison of the observational data stack and the Gaussian
absorption model of the Monte Carlo simulations.
n the figure, the Gaussian absorption model is
broadened by the mean redshift error by adding it in quadrature to the dispersion in the model.
The resulting stack of mock spectra has a $1\sigma$ dispersion of $282\,{\rm km\,s^{-1}}$,
about 2 times the modeling error away from the intrinsic dispersion in the \ion{C}{2} mean stack
for the full sample, and about 2 times the modeling error away from the intrinsic dispersion in the
stack of the sub-sample with [\ion{O}{3}] redshifts as well.
One may also consider whether absorbers in Hubble flow and cosmological distances show peculiar
velocities that are typical for quasar-mass halos, and adopt a different $\sigma_{\rm rms}$ accordingly.
The best-studied coeval, more typical star-forming
galaxies are the Lyman-break galaxies. Their clustering strength implies a characteristic halo
mass of $M_{\rm halo}\approx10^{11.6}\,M/{\rm M_\odot}$ \citep{Bielby+13,Malkan+17}, corresponding to
$\sigma_{\rm rms}=114\,{\rm km\,s^{-1}}$.
If we adopt this smaller $\sigma_{\rm rms}$ insead,for absorbers outside the loosely
defined quasar CGM boundary, at $\gtrsim300$\,kpc, the stack of mock spectra will have a $1\sigma$
dispersion of $247\,{\rm km\,s^{-1}}$. This is again within modeling errors of the observed intrinsic
dispersion.
We also test for the
sensitivity of this measured dispersion to the correlation functions adopted. The QPQ6 clustering
analysis is performed on only the strongest absorber near $z_{\rm fg}$, and a low $R_\perp$
sightline may in fact intercept more than one optically thick absorber. We double the number of
absorbers for each mock sightline, and find the measured dispersion only increases by several
${\rm km\,s^{-1}}$. Thus, before the putative asymmetry is confirmed, the hypothesis that the
observed velocity width is only produced by a
combination of gravitational motions and Hubble flows cannot be ruled out. Extra dynamical
processes are not necessary to explain the large velocity fields.

Although we acknowledge the possibility that our Gaussian model does not capture all the powers
at extreme velocities, we also caution that occasional large kinematic offsets do not make a strong
case against gravitational motions. Firstly, quasars occasionally inhabit extremely overdense environments
such as protoclusters \citep[e.g.,][]{Hennawi+15}. Secondly, the probability of intercepting a random,
unassociated absorber is boosted by large-scale clustering. Hence occasional extreme velocities would not
necessitate outflows. Our preferred kinematic measure is thus the dispersion in the average absorption.

In Figure~\ref{fig:contour}, We show the probability distributions of the degenerate parameters
$W_{\rm CII}$ and $M_{\rm halo}$ in recovering the intrinsic width of the absorption profile. We
require that the amplitude of the absorption is reproduced within 3 times its modeling error, and
mark contours for points in ($W_{\rm CII}$, $M_{\rm halo}$)-space that produce an absorption
profile of width within 1, 2, and 3 times the modeling error of the observed intrinsic width.
From the figure, a higher $W_{\rm CII}$ means the $M_{\rm halo}$ that will reproduce the observed
width is higher. If there are no extra dynamical processes, the intrinsic velocity width
corresponding to typical QPQ halo mass is contained within the 2$\sigma$ contour.

\subsection{Interpretation of the asymmetric absorption}
\label{sec:significance_+ve}

\begin{figure}
\includegraphics[width=2.5in]{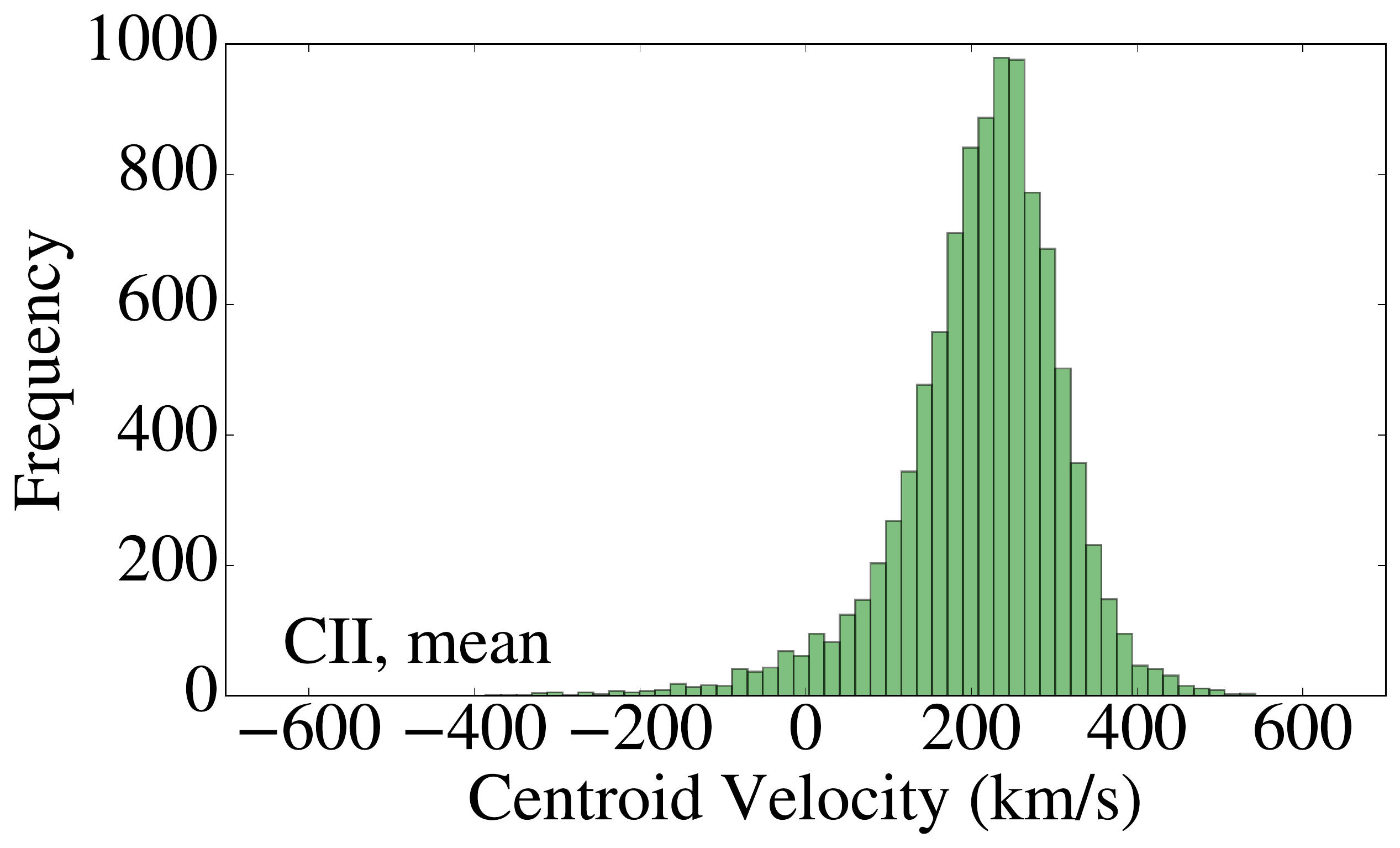}
\caption{Histogram of the absorption centroids of 10000 bootstrap realizations of the data sample
for the \ion{C}{2} mean stack. 97\% of the centroids are positively offset from $z_{\rm fg}$.}
\label{fig:histogram_cen}
\end{figure}

We quote the standard deviation in the bootstrap
realizations to be the scatter in the centroids of the data. The scatters
$\approx100\,{\rm km\,s^{-1}}$ are comparable to the measured offsets, indicating large intrinsic
variation in quasar CGM environments (see also QPQ8). In Figure~\ref{fig:histogram_cen}, we show
the distribution of the absorption centroids from bootstrapping on the \ion{C}{2} mean stack. We
find that 97\% of the centroids are at positive velocities. We place a generous $3\sigma$ upper
limit to the small offset of the centroid from $z_{\rm fg}$ for the \ion{C}{2} mean absorption at
$\delta v<+526\,{\rm km\,s^{-1}}$. Given the large intrinsic scatter, we do not attempt to explore
whether there exists relative asymmetry among the \ion{C}{2}, \ion{C}{4}, and \ion{Mg}{2}
absorption.

One may ask whether absorption from \ion{C}{2*}\,1335 may bias the measurement of the
\ion{C}{2}\,1334 velocity centroid. In the stacks at \ion{C}{2}\,1334, absorption from \ion{C}{2*}\,1335
would manifest as a component redshifted by $+264\,{\rm km\,s^{-1}}$.
In the higher resolution data from QPQ8, we identify nine \ion{C}{2}-bearing subsystems where absorption
from \ion{C}{2*}\,1335 is not blended with \ion{C}{2}\,1334.
Their mean \ion{C}{2} to \ion{C}{2*} equivalent width ratio is 0.2.
Among these nine subsystems, only one,
labeled J1420+1603F in QPQ8, has a \ion{C}{2*}\,1335 equivalent width comparable to \ion{C}{2}\,1334.
We simulate that absorption from \ion{C}{2*} may only bias the \ion{C}{2} centroid by
$\approx+40\,{\rm km\,s^{-1}}$.
Hence, contamination from \ion{C}{2*}\,1335 could not be responsible for the observed redshift in
the centroid.

One may also ask whether the measured positive offsets come from systematic bias in redshift
measurements due to the Baldwin effect \citep{Baldwin77}. \cite{Shen16} reported that, the
[\ion{O}{3}] emission of $z\sim2$ quasars is more asymmetric and weaker than that in typically
less luminous low-$z$ quasars. To test for this potential source of bias, we create another mean
stack at \ion{C}{2}\,1334 by replacing the [\ion{O}{3}] redshifts by a redshift measured from the
more symmetric \ion{Mg}{2} or H$\alpha$ emission when available. We are able to replace for 11 out
of the 15 systems with [\ion{O}{3}] redshifts in the original sample. The new stack is similiar in
velocity structure and again shows a positive offset $\approx+303\,{\rm km\,s^{-1}}$. We thus
conclude that our algorithm for measuring redshifts is not severely biased by the blue wing of the
[\ion{O}{3}] emission-line.

\begin{figure}
\includegraphics[width=2.5in]{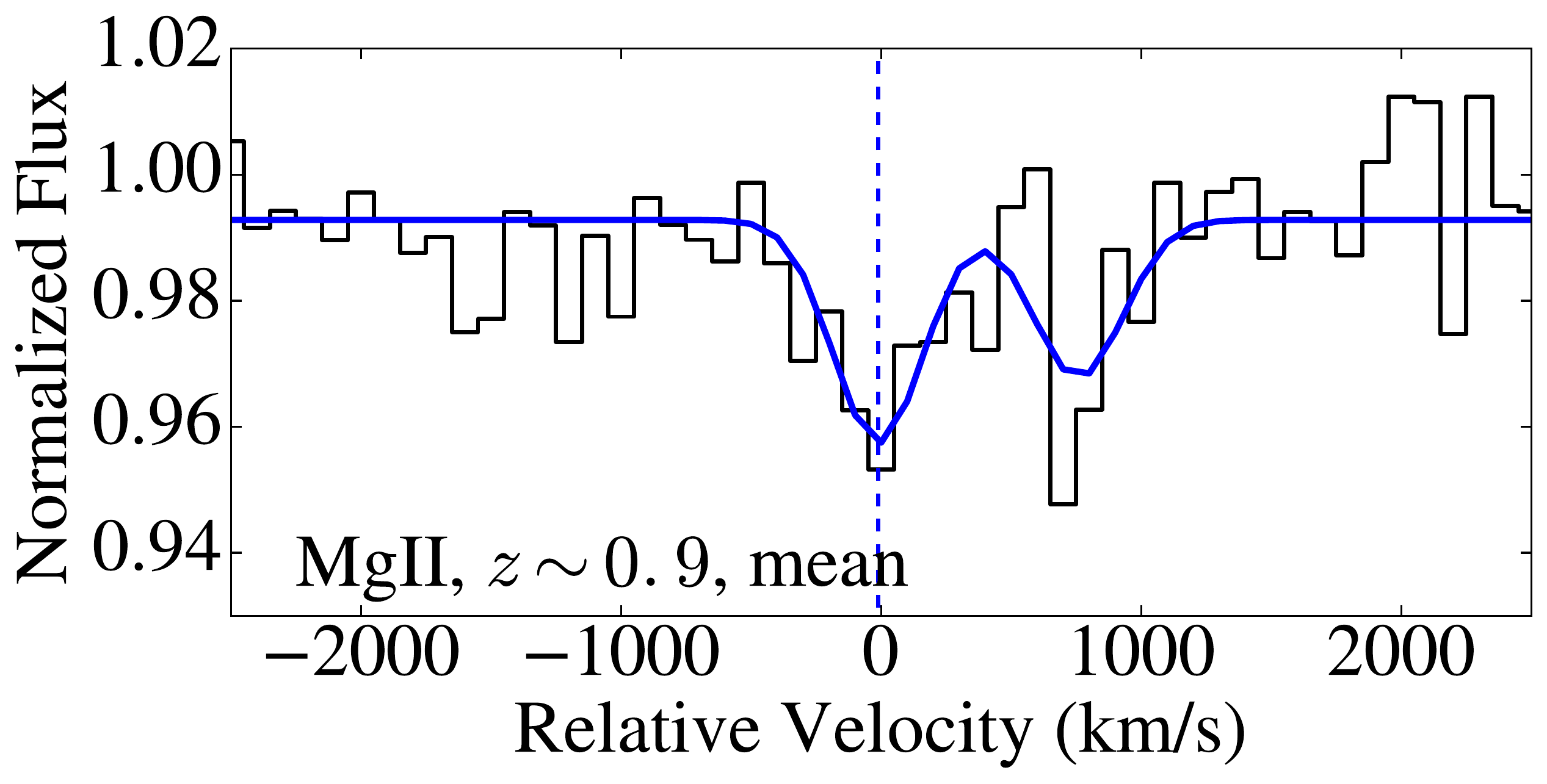}
\caption{Mean stack at \ion{Mg}{2}\,2796, for a lower redshift sample of quasar pairs at
$z\sim0.9$. Line-style coding is the same as in Figure~\ref{fig:stacks_and_fits}. The centroid is
approximately at $0\,{\rm \,km\,s^{-1}}$, and the absorption is weaker than the main QPQ9 sample.
}
\label{fig:stack_z1}
\end{figure}

Motivated by the study of \ion{Mg}{2} absorbers surrounding $z\sim1$ quasars by \cite{Johnson+15},
we also generate a mean-stacked spectrum for \ion{Mg}{2}\,2796 for lower redshift quasar pairs.
We select quasars with $0.4<z_{\rm fg}<1.6$ and use the same other selection criteria as the main
QPQ9 sample. The quasar pairs are selected from the igmspec database, with $z_{\rm fg}$ measured
by \cite{HewettWild10}. For quasars with $z_{\rm fg}<0.84$, as redshift determination is dominated
by [\ion{O}{3}] emission, a shift of $+48\,{\rm km\,s^{-1}}$ is applied to bring the emission-line
redshift to the systemic. For quasars with $0.84<z_{\rm fg}<1.6$, as redshift determination is
dominated by \ion{Mg}{2} emission, a shift of $+57\,{\rm km\,s^{-1}}$ is applied. There are 233
pairs selected, with a median $z_{\rm fg}$ of 0.90 and a median $R_\perp$ of 208\,kpc. We present
the mean stack in Figure~\ref{fig:stack_z1}. The absorption is weaker than the $z\sim2$ main QPQ9
sample. Gaussian absorption models fitted to the stack recover a centroid of
$-11\pm379\,{\rm km\,s^{-1}}$ and a dispersion of $172\,{\rm km\,s^{-1}}$. The average offset
from $0\,{\rm km\,s^{-1}}$ is much smaller than the offsets in the $z\sim2$ sample.

Since the large scatter in the centroids represents intrinsic variation rather than redshift
errors, and the \ion{Mg}{2} stack for lower redshift suggests a different centroid, we consider
that
systematic biases are unlikely to explain the asymmetry signal in the $z\sim2$ sample.
In the Discussion section, we discuss two possible explanations for the asymmetry.

\section{Discussion}
\label{sec:discussion}


\sidecaptionvpos{figure}{c}
\begin{SCfigure*}[10][!th]
\includegraphics[height=0.49\textheight]{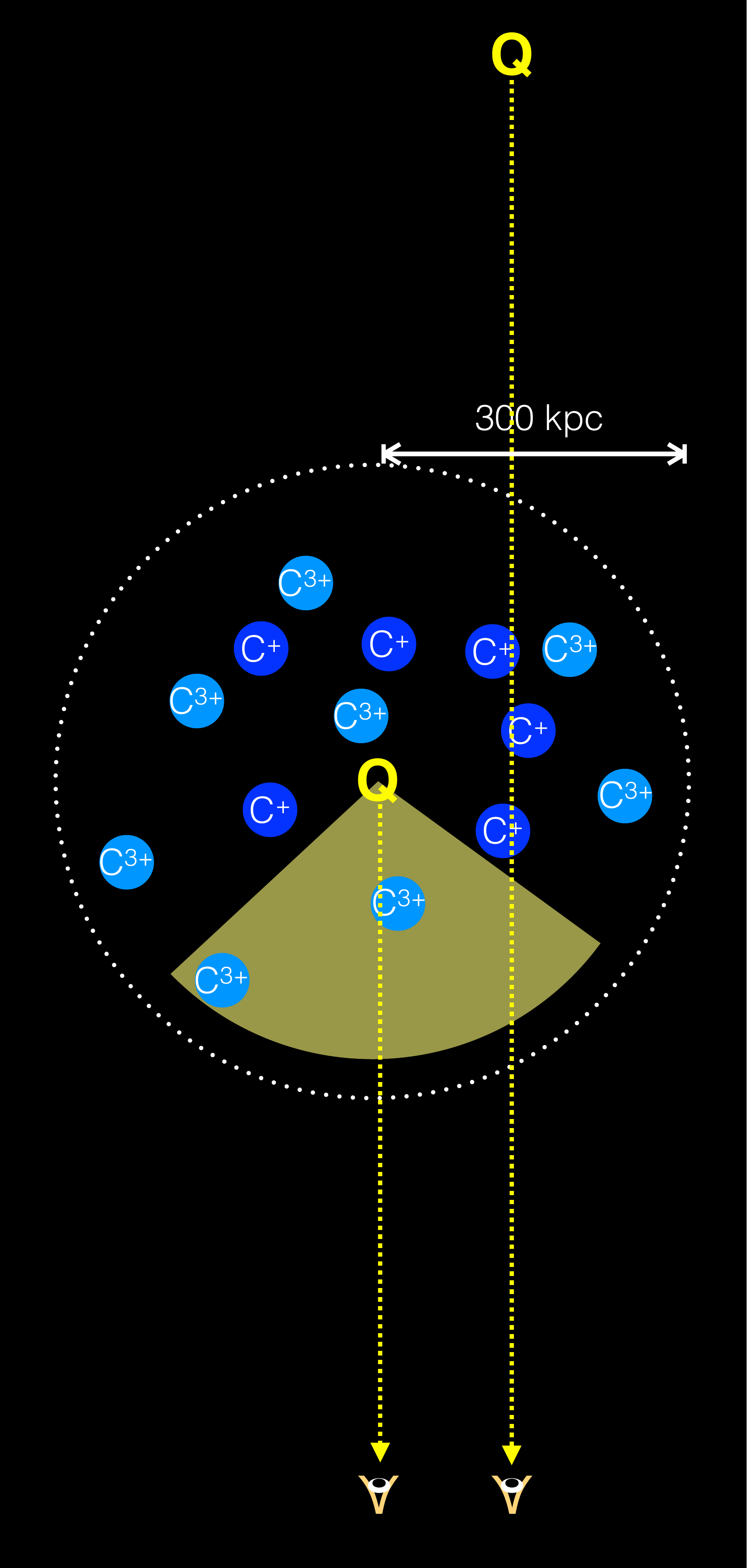}
\caption{A cartoon showing a unipolar quasar. The gas observed in low- to intermediate-ion
absorption preferentially lies behind the quasar, and is shadowed from the ionizing radiation.}
\label{fig:monopolar}
\end{SCfigure*}
\sidecaptionvpos{figure}{c}
\begin{SCfigure*}[10][!bh]
\includegraphics[height=0.49\textheight]{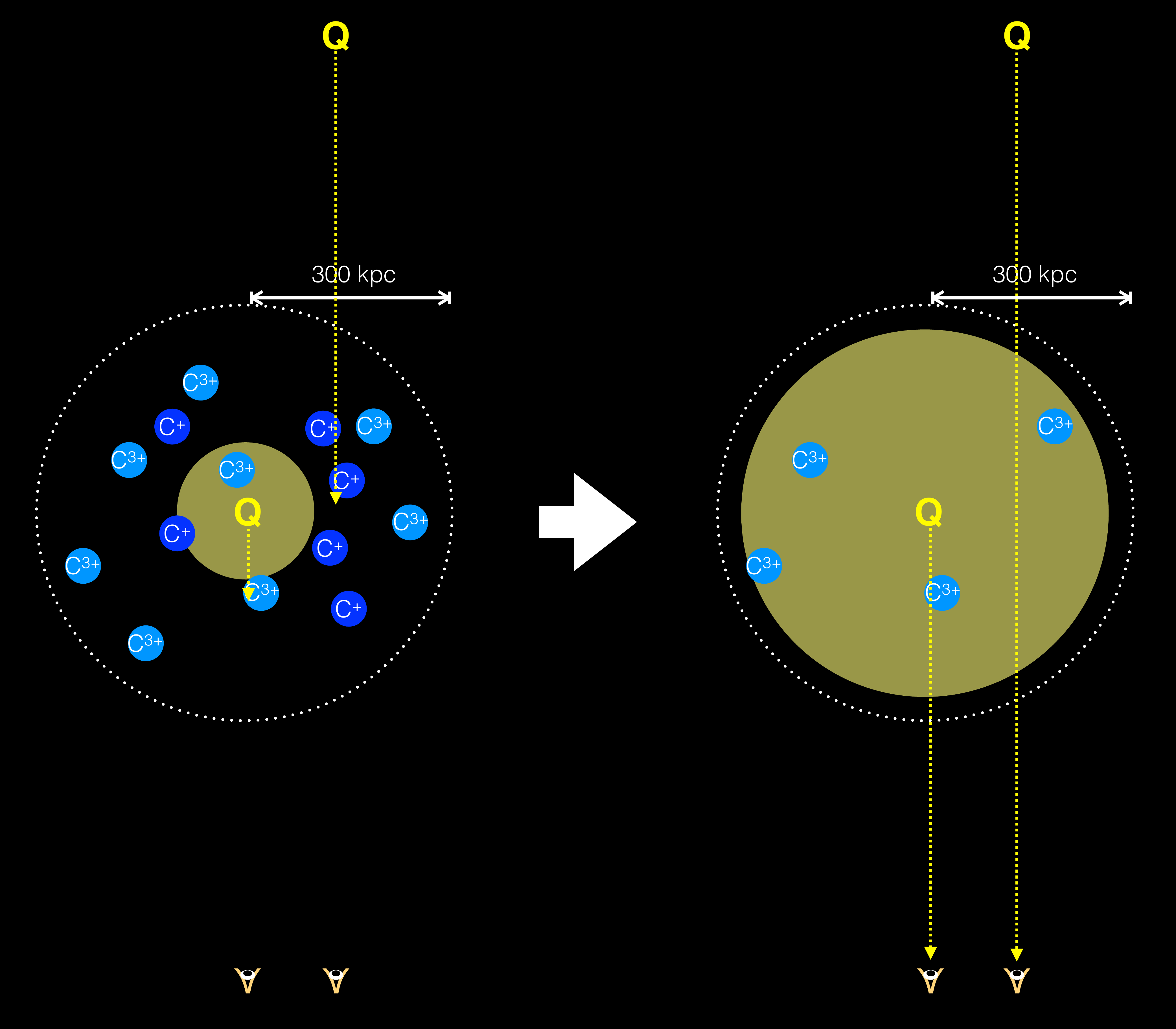}
\caption{A cartoon showing the finite lifetime of quasar episodes as an explanation to the
asymmetric absorption. The setup on the left shows that the foreground quasar has not been shining
long enough for its ionizing radiation to reach the gas behind it, when the light from the
background quasar reaches. The setup on the right shows the scenario after an amount of time
comparable to the light travelling time across CGM scale. Gas in front of the foreground quasar
has been ionized, by the time the light from the background quasar reaches.}
\label{fig:lighttravel}
\end{SCfigure*}

At low redshift, both symmetric and asymmetric ionization cones around quasars and AGNs are
observed. In the extended emission-line region of 4C37.43, most of the [\ion{O}{3}] emission is
blueshifted \citep{FuStockton07}, but there are counter examples in less extended sources in
\cite{FuStockton09}. Recently, there has been Fabry-Perot interferometric data for less extended
narrow-line regions in more nearby sources \citep{Keel+15,Keel+17}, which show mostly symmetric
velocity fields and gas distributions.

In the following, we explore two possible explanations for the non-dynamical processes that
provides the putative asymmetry.
The explanations arise from a transverse proximity effect (QPQ4), which is the suppression of
opacity in background sightlines passing close to foreground quasars.

One possibility is an asymmetric radiation field that preferentially ionizes the gas moving toward
the observer, where the quasar is known to shine. Alternatively, the asymmetric radiation field
may preferentially ionizes the gas at smaller Hubble velocity than the quasar. In
Figure~\ref{fig:monopolar}, we show a cartoon of a quasar that is blocked in the direction
pointing away from the observer. The gas observed in absorption preferentially lies behind the
quasar.
\cite{Roos+15} and \cite{GaborBournaud14} performed simulations of a high-redshift disk galaxy
including thermal AGN feedback and calculated radiative transfer in post-processing. They found
the ionization radiation is typically asymmetric, due to either a dense clump that lies on one
edge of the black hole or the black hole's location being slightly above the disk.
\cite{FaucherGiguere+16} represents the only simulation so far that is able to reproduce the
substantial amount of cool gas in quasar-mass haloes. We eagerly await their group to compare the
fraction of gas in inflows and outflows.

Another possible explanation arises from the finite lifetime of quasar episodes.
Figure~\ref{fig:lighttravel} presents a cartoon for a light travel time argument. The light from
the background quasar may arrive at the gas behind
the foreground quasar before the ionizing radiation from the foreground quasar arrives.


The first explanation above to asymmetric
absorption requires that the quasars emit their ionizing radiation anisotropically, while the
second explanation requires the quasars emit their ionizing radiation intermittently. Both
explanations will require the line-of-sight motions of the gas to be not in a net inflow.
Were the gas flowing into the galaxy instead, the velocity centroid would be negative.
Under both scenarios, lows ions should be redshifted while high ions should be blueshifted.
Using Cloud photoionization models, we find that, for the typical quasar luminosity of our sample,
a CGM gas that is directly illuminated is highly ionized and shows only marginally detectable
\ion{C}{4} absorption. The redshifted \ion{C}{4} may be regarded as an intermediate ion, and a
blueshifted absorption signal needs to be searched in a higher ion.

We test the case of Hubble flows versus galactic-scale outflows in producing the putative
asymmetry.
Anistropic emission is degenerate with intermittent emission in their asymmetric
light-echo in the observer's frame. We first consider the scenario where
the quasar's lifetime is infinite, and the observed ansymmetric absorption is only produced by
anisotropic emission.
We try implementing an arbitrary quasar opening angle in our Monte Carlo simulations to
reproduce an absorption centroid that is redshifted from the systemic. In QPQ6, we argued that
the observed anisotropic clustering of optically thick systems around quasars demands that a quasar's
radiation field must affect optically thick systems on Mpc scales. Hence in this test, we set the
incidence of absorbers happening within the quasar opening angle to be zero, for absorbers at all
disances from the quasar. In the case without
outflows, even if we set the opening angle to be $180^\circ$, i.e.\ absorption only happens at
positive velocities, the mean absorption centroid would merely reach $\approx+85\,{\rm km\,s^{-1}}$.
Hence, the observed $\delta v\approx+200\,{\rm km\,s^{-1}}$ shift in the mean absorption cannot be
produced by asymmetric distribution in Hubble velocities alone. This suggests the presence of an
outflow component to account for extra asymmetry in line-of-sight velocities.
In order to produce the observed intrinsic dispersion and centroid of the mean absorption by
\ion{C}{2}, we must add a radial outflow speed of $\approx450\,{\rm km\,s^{-1}}$ to the absorbers
and a unipolar quasar opening angle of $\approx180^\circ$.
Next, we consider another scenario where the quasar is isotropic, and the asymmetric absorption
is only produced by short episodic lifetime. We further make the assumption the luminosity is
constant during a quasar episode. In the quasar's rest frame, the light echo
would be observed as a spherical region with the quasar at the origin. In the observer's frame, due
to finite light travel time, the light-echo would not be spherically symmetric. The
light-echo from the quasar at its luminosity $t$\,yr earlier traces a paraboloid
with the quasar at the focus and the vertex $t/2$\,ly behind it \citep{Adelberger04,VisbalCroft08}.
We find that, to reproduce the observation, the CGM absorbers need to have a radial outflow speed
$\approx420\,{\rm km\,s^{-1}}$ and the quasar must have shined for $\approx0.4\times10^6$\,yr.
For the anisotropy-only scenario, the opening angle deduced is rather large compared to literature
findings, which give $30^\circ\textendash90^\circ$ \citep[e.g.,][]{TrainorSteidel13,Borisova+16}.
For the intermittence-only scenario, if the quasars are on
averaged observed near the middle of the episode, the lifetime deduced is somewhat small compared
to other existing constraints from observations and simulations, which give
$10^6\textendash10^8$\,yr \citep[e.g.,][]{Martini04,Hopkins+05}.
We thus speculate that the asymmetric absorption is the result of a combination of anisotropic
and intermittent emission.

We note that, \cite{Turner+17} conclude that the clustering of low to intermediate ions around
Lyman-break galaxies in velocity space is
most consistent with gas that is inflowing on average. The different conclusion from our analysis
may originate from the higher masses of our systems, the presence of quasar-driven ouflows
\citep[e.g.,][]{Greene+12}, and/or starburst-driven outflows that are correlated with the presence
of quasars \citep[e.g.,][]{Barthel+17}.

Motivated by the asymmetry found in metal ion absorption in the CGM using precise $z_{\rm fg}$
measurements, and the asymmetry found by \cite{KirkmanTytler08} in \ion{H}{1} on larger
scales, we are assembling a sample of quasar pairs with precise $z_{\rm fg}$ measurements to study
this asymmetry in \ion{H}{1} (J.\ F.\ Hennawi et al.\ 2018, in preparation). In conclusion, we
observe large and positively skewed velocity fields in absorption, of metal ions
in the CGM of $z\sim2$ massive galaxies hosting quasars.
We argue that, the observation of large velocity fields alone can be accounted for by
gravitational motions and Hubble flows and does not necessitate outflows. However, we argue that the
positive skew suggests the detected gas is in outflow on average, and the quasars shine
preferentially toward the observer and/or intermittently.

\acknowledgements
MWL is immensely grateful to the referee's constructive comments, which help strengthen the
paper.
JXP and MWL acknowledge support from the National Science Foundation (NSF) grants AST-1010004 and
AST-1412981. The authors gratefully acknowledge the support which enabled these observations at
the Keck, Gemini, Large Binocular Telescope, Very Large Telescope, Las Campanas, and Palomar
Observatories. The authors acknowledge the sharing of private data by Yue Shen. MWL thanks Hai Fu
for a discussion on extended emission-line regions, T.-K. Chan for a discussion on CGM
simulations, and Chi Po Choi and Yat Tin Chow for discussions on statistics.

\appendix

\section{Line-of-sight absorption}
\label{sec:appendix}

\begin{figure}
\includegraphics[width=3.4in]{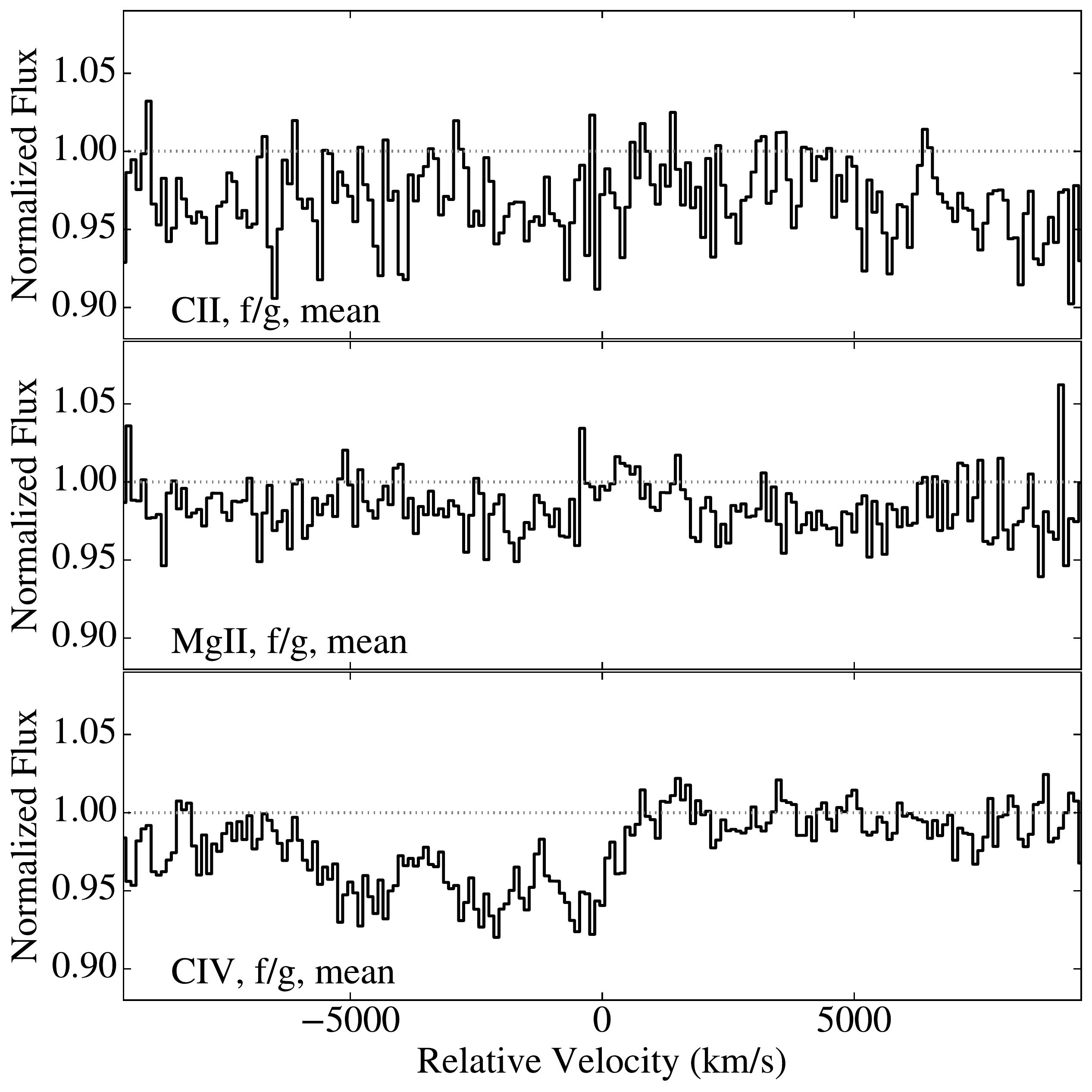}
\caption{Mean stacks of the foreground quasar spectra at \ion{C}{2}\,1334, \ion{C}{4}\,1548, and
\ion{Mg}{2}\,2796 for the QPQ9 sample. For \ion{C}{2} and \ion{Mg}{2}, the mean absorption is
weaker than that in the background stacks, and there is no evidence for an excess at $z_{\rm fg}$.
The stack for \ion{C}{4}, which includes line-of-sight absorbers at all distances, shows a large,
blueshifted mean velocity field.
}
\label{fig:stacks_fg}
\end{figure}

In the previous QPQ papers, we argued that optically thick absorbers in the vicinity of quasars
are distributed anisotropically. We now have the means to show this anisotropic clustering
explicitly. Given that the techniques for stacking spectra are established, it is straightforward
to apply the same techniques to stack the foreground quasar spectra. In
Figure~\ref{fig:stacks_fg}, we present mean stacks of foreground quasar spectra for
the QPQ9 sample. We require that the spectra survive a S/N cut of 5.5 per rest-frame \AA\ at
\ion{C}{2}\,1334, \ion{C}{4}\,1548, or \ion{Mg}{2}\,2796 at $z_{\rm fg}$. In contrast to the
large equivalent widths exhibited in the stacks of background spectra, \ion{C}{2} and \ion{Mg}{2}
mean absorption along the line-of-sight to the foreground quasars is weaker, and an excess at
$z_{\rm fg}$ is absent. This supports a scenario where the ionizing radiation of the foreground
quasars are anisotropic and/or intermittent. For \ion{C}{4}\,1548, this stack of all line-of-sight
absorbers, which include absorbers intrinsic to and far away from the quasar, shows a large,
blueshifted velocity field. This excess \ion{C}{4} absorption has been well studied as narrow
associated absorption line systems \citep[e.g.,][]{Wild+08}.
We note that, in the $z\sim4$ sample of narrow associated absorbers analyzed by \cite{Perrotta+16} 
and Perrotta et al.\ (2018, in preparation), a higher ionization state is similarly found along 
the line-of-sight to quasars compared to across the line-of-sight. In particular, an excess of 
\ion{N}{5} absorption is found in proximity to the host quasars. 


\end{document}